\documentclass[journal]{IEEEtran}

\ifCLASSINFOpdf

\else

\fi

\usepackage[OT1]{fontenc} 
\usepackage{cite}
\usepackage{amsmath,amssymb,amsfonts}
\usepackage{algorithmic}
\usepackage{graphicx}
\usepackage{textcomp}
\usepackage{xcolor}
\usepackage{subfigure}
\usepackage{verbatim} 
\newcommand{\tabincell}[2]{\begin{tabular}{@{}#1@{}}#2\end{tabular}} 
\usepackage{multirow} 
\usepackage[OT1]{fontenc} 
\usepackage{diagbox}
\usepackage{url}
\usepackage{color}
\usepackage{enumitem} 
\usepackage[ruled,linesnumbered]{algorithm2e}
\usepackage{threeparttable}

\begin{document}
%
\title{Subjective Quality Database and Objective Study of Compressed Point Clouds with 6DoF Head-mounted Display}

%
%
%

\author{Xinju~Wu,
Yun~Zhang,~\IEEEmembership{Senior Member,~IEEE,}
Chunling~Fan,
Junhui~Hou,~\IEEEmembership{Senior Member,~IEEE,}
and Sam~Kwong,~\IEEEmembership{Fellow,~IEEE}
\thanks{This work was supported in part by Guangdong International Science and Technology Cooperative Research Project under Grant 2018A050506063, in part by Shenzhen Science and Technology Program under Grant JCYJ20180507183823045 and JCYJ20200109110410133, in part by Membership of Youth Innovation Promotion Association, Chinese Academy of Sciences under Grant 2018392. \textit{(Corresponding author: Yun Zhang)}}
\thanks{X. Wu is with the Shenzhen Institute of Advanced Technology, Chinese Academy of Sciences, Shenzhen 518055, China, and also with Shenzhen College of Advanced Technology, University of Chinese Academy of Sciences, Shenzhen 518055, China (e-mail: xj.wu1@siat.ac.cn).}
\thanks{Y. Zhang is with the Shenzhen Institute of Advanced Technology, Chinese Academy of Sciences, Shenzhen 518055, China (e-mail: yun.zhang@siat.ac.cn).}
\thanks{C. Fan is with the School of Electronic and Communication Engineering, Shenzhen Polytechnic, Shenzhen 518055, China (e-mail: fan\_chunling@126.com).}
\thanks{J. Hou and S. Kwong are with the Department of Computer Science, City University of Hong Kong, Hong Kong, and also with the City University of Hong Kong Shenzhen Institute, Shenzhen 518057, China (e-mail: jh.hou@cityu.edu.hk; cssamk@cityu.edu.hk).}
}
%
%

\markboth{IEEE TRANSACTIONS ON CIRCUITS AND SYSTEMS FOR VIDEO TECHNOLOGY}%
{Shell \MakeLowercase{\textit{et al.}}: Bare Demo of IEEEtran.cls for IEEE Journals}
%



\makeatletter
\def\ps@IEEEtitlepagestyle{%
  \def\@oddfoot{\mycopyrightnotice}%
  \def\@oddhead{\hbox{}\@IEEEheaderstyle\leftmark\hfil\thepage}\relax
  \def\@evenhead{\@IEEEheaderstyle\thepage\hfil\leftmark\hbox{}}\relax
  \def\@evenfoot{}%
}
\def\mycopyrightnotice{%
  \begin{minipage}{\textwidth}
  \centering \scriptsize
  Copyright~\copyright~2021 IEEE. Personal use of this material is permitted. However, permission to use this material for any other purposes must be obtained from the IEEE by sending an email to pubs-permissions@ieee.org.
  \end{minipage}
}
\makeatother

\maketitle

\begin{abstract}
In this paper, we focus on subjective and objective Point Cloud Quality Assessment (PCQA) in an immersive environment and study the effect of geometry and texture attributes in compression distortion. 
Using a Head-Mounted Display (HMD) with six degrees of freedom, we establish a subjective PCQA database, named SIAT Point Cloud Quality Database (SIAT-PCQD). Our database consists of 340 distorted point clouds compressed by the MPEG point cloud encoder with the combination of 20 sequences and 17 pairs of geometry and texture quantization parameters. 
The impact of distorted geometry and texture attributes is further discussed in this paper. 
Then, we propose two projection-based objective quality evaluation methods, i.e., a weighted view projection based model and a patch projection based model. 
Our subjective database and findings can be used in point cloud processing, transmission, and coding, especially for virtual reality applications. The subjective dataset\footnote{https://dx.doi.org/10.21227/ad8d-7r28}~\footnote{http://codec.siat.ac.cn/video\_download\_siat-pcqd.html} has been released in the public repository.
\end{abstract}


%
\IEEEpeerreviewmaketitle

\section{Introduction}
%
%
%
%
\IEEEPARstart{I}{N} recent years, the significant advance of Two-Dimensional (2D) video is the improvement of resolution, which has evolved from Standard Definition (SD), High Definition (HD) to Ultra High Definition (UHD). Visual information perceived by humans at the moment has been enriched with the progress of resolution. 
Meanwhile, people gradually attach importance to the experience of watching videos, especially preferring active interaction. Extended Reality (XR) technology, summarizing Virtual Reality (VR), Augmented Reality (AR), and Mixed Reality (MR) technologies, has attracted much attention with the emergence of Head-Mounted Display (HMD). 
In XR, degree of freedom refers to the head movement in space, with three corresponding to rotation movement, and the other three corresponding to translation movement. However, 360-degree videos only support Three Degrees of Freedom (3DoF), tracking users’ rotational motion but not translational movement. To keep pace with the consumption of visual media, some novel concepts of immersive media came out, like 3DoF+, i.e., enabling additional limited translational movements, and Six Degrees of Freedom (6DoF), allowing rotational motion as well as translational motion.
In the VR environment with 6DoF, observers utilize eye and head movement to explore scenes,  utterly different from traditional 2D viewing. 
Many types of data are applicable for 6DoF, such as simple proxy geometry, voxels, and point clouds.

A point cloud is a collection of Three-Dimensional (3D) points in 3D space without space connections or ordering relations~\cite{schwarz_emerging_2019}. Each point in a point cloud consists of a geometry attribute, i.e., 3D position $(x, y, z)$, and other attributes like color, reflectivity, and opacity denoted by vectors. According to the Point Cloud Compression (PCC) group of MPEG, point clouds can be categorized as static, dynamic, and dynamically acquired point clouds~\cite{schwarz_emerging_2019}. Similar to the relationship between videos and images, a dynamic point cloud recognizes each static point cloud as a frame, showing the movement of a 3D object or a scene going after the temporal variation. Targeting XR applications, static or dynamic point clouds can be captured in a studio full of high-speed cameras, especially for contents like people and objects. Targeting applications like autonomous driving, dynamically acquired point clouds are mainly obtained by LIDAR sensors on the top of a moving vehicle, enabling dynamic environment perception in robot navigation. For superior perceptual experience, point clouds are desired to be captured densely with high precision. Unlike meshes, point clouds have no spatial connectivity nor ordering relations, excluding the concepts of edges, faces, or polygons. The massive number of points and the inherent characteristics of disjunction and disorder pose challenges for point cloud processing and compression.


For storage and transmission of point clouds, efficient compression frameworks were researched by scholars. In~\cite{mekuria_design_2016}, a time-varying point cloud codec was first proposed in a 3D tele-immersive system and later regarded as the anchor of MPEG PCC standards. The emerging PCC is composed of two classes for compression for different categories of point clouds. One is Video-based PCC (V-PCC) targeting dynamic point clouds, and the other is Geometry-based PCC (G-PCC). G-PCC is the combination of Surface PCC (S-PCC) for static point clouds and LIDAR PCC (L-PCC) for dynamically acquired point clouds. Nowadays, PCC~\cite{schwarz_emerging_2019
} is a trending and intriguing research topic. 

Both processing and compression may induce kinds of distortions for point clouds, including changes in the number of points and values of positions and colors. 
The degradation of the contents might influence users' perception. Therefore, subjective and objective assessment is the pointed and valid method to reflect the quality of point clouds. Nowadays, point cloud evaluation is still a sophisticated and challenging problem involved with excessive factors such as degradation, rendering methods, display equipment, evaluation methodologies, quality of sources and so forth. 

A normative and effective framework for Point Cloud Quality Assessment (PCQA) is necessitous but has not been explored fully yet.
Learning from Image Quality Assessment (IQA)
 and Video Quality Assessment (VQA)
, many papers for PCQA came out since 2017. However, the early works~\cite{alexiou_towards_2017, javaheri_subjective_2017, alexiou_performance_2017, alexiou_impact_2018, alexious_point_2018, alexiou_point_2018-2} only focused on colorless point clouds and simple types of degradation, while point clouds with color and codecs like V-PCC are the prevailing trend. Also, some works utilized a virtual camera around point clouds to render videos, and those videos were then displayed in a 2D screen for subjects to evaluate, which lacked human interaction.

In this work, we concentrate on subjective PCQA in an immersive 6DoF VR environment to study the effect of geometry and texture attributes in compression distortion. First, contents like human figures and inanimate objects are selected, showing the variety of test sequences. 
A total of 17 different combinations of geometry and texture Quantization Parameters (QP) with 20 sequences are used to create 340 distorted point clouds by the V-PCC codec. 
Our subjective database, named SIAT Point Cloud Quality Database (SIAT-PCQD)~\cite{siat-pcqd}, is available on public repository. 
Second, we analyze essential factors in the visual quality of point clouds in detail, including geometry and texture attributes from the perspective of different QP levels and types of sequences.
Finally, we propose two projection-based objective quality evaluation methods, i.e., a weighted view projection based model and a patch projection based model.

The remainder of our paper is organized as follows. Section~\ref{section2} reviews the related work in subjective and objective point cloud evaluation. Section~\ref{section3} details our subjective quality evaluation test, including procedures like data preparation, rendering techniques, equipment and environment, and evaluation methodology. Section~\ref{section4} shows data processing and the results of our subjective experiment. 
In Section~\ref{section5}, we propose two projection-based objective quality evaluation methods, and the performance evaluation will be shown in Section~\ref{section6}.  
Finally, the conclusion of our work and the challenges of PCQA are outlined in Section \ref{section7}.

\section{Related Works}\label{section2}
In this section, related works in point cloud evaluation are described in the aspects of subjective and objective assessment.

\subsection{Subjective Point Cloud Quality Assessment} \label{section2: subjective PCQA}

\begin{table*}
\caption{Summary of subjective PCQA databases.}
\begin{center}
\resizebox{\textwidth}{!}{
\begin{tabular}{l|c|c|c|l|l|c|c|c}
\hline
Method & Sequence set & \tabincell{l}{Status of\\ sequences} & \tabincell{l}{Colored} & Degradation & \tabincell{l}{\#Distorted point clouds\\$seq\times dist\times rate$} & Display & \tabincell{l}{Inter-\\action} & Methodology \\
\hline

Alexiou \textit{et al.}~\cite{alexiou_towards_2017} & objects & \multirow{3}{*}{static} & \texttimes &  \tabincell{l}{Gaussian noise, octree-pruning} & $5\times 2\times 4=40$ & AR & \multirow{3}{*}{\checkmark} &  DSIS \\
\cline{5-6}
Torlig \textit{et al.}~\cite{torlig_novel_2018} & objects, humans &   & \checkmark  & octree-based compression \&JPEG  & $7\times 9=63$ & 2D monitor & & DSIS \\
\cline{5-6}
Alexiou \textit{et al.}~\cite{alexiou_comprehensive_2019}   & objects, humans &  & \checkmark &  V-PCC, G-PCC   & $9\times (5+6)=99$ &  2D monitor & &   DSIS \\
\hline

Javaheri \textit{et al.}~\cite{javaheri_subjective_2017} & \multirow{5}{*}{objects} & \multirow{5}{*}{static} & \multirow{5}{*}{\texttimes} &  \tabincell{l}{2 outerlier removal algorithms,\\3 denoising algorithms}   & $4\times 5\times 3=60$  & 2D monitor & \multirow{5}{*}{\texttimes} &   DSIS \\
\cline{5-6}
Alexiou and Ebrahimi~\cite{alexiou_performance_2017} &  &  &   & \multirow{2}{*}{Gaussian noise, octree-pruning} & $5\times 2\times 4=40$& 2D monitor &  &  DSIS, ACR\\
Alexiou and Ebrahimi~\cite{alexiou_impact_2018} &  &  &  &    & $5\times 2\times 4=40$ & 2D monitor & &  DSIS\\
\cline{5-6}
Alexiou \textit{et al.}~\cite{alexious_point_2018} &  & &   & \multirow{2}{*}{octree-pruning}  & $7\times 1\times 4=28$ & 2D monitor & & DSIS \\
Alexiou \textit{et al.}~\cite{alexiou_point_2018-2} &   &   &&   & $7\times 1\times 4=28$ & 2D/3D monitor  & &  DSIS\\
\hline

Zhang \textit{et al.}~\cite{zhang_subjective_2014}  &  objects &static & \multirow{17}{*}{\checkmark} &  \tabincell{l}{down-sampling,\\ geometry noise, color noise}  & $1\times (6+7+12)=25$ & \multirow{17}{*}{2D monitor} &  \multirow{17}{*}{\texttimes} & - \\
\cline{5-6}
Javaheri \textit{et al.}~\cite{javaheri_subjective_2017-1} &  objects, humans & static & & \tabincell{l}{octree-pruning,\\ graph-based compression}  & $6\times 2\times 3=36$ &  & & DSIS \\
\cline{5-6}
da Silva Cruz \textit{et al.}~\cite{da_silva_cruz_point_2019} & \tabincell{c}{objects, humans,\\ scenes} & static & &\tabincell{l}{octree-pruning,\\ projection-based encoder}  & $8\times 2\times 3=48$  &   & & DSIS\\
\cline{5-6}
SJTU-PCQA~\cite{9238424} & objects, humans & static & & \tabincell{l}{octree-based compression, color noise, \\geometry noise, scaling} & $10\times 7\times 6=420$& & & ACR \\
\cline{5-6}
vsenseVVDB~\cite{zerman_subjective_2019} & humans & dynamic &  &  down-sampling, V-PCC  & $2\times 2\times 4=16$ & &  &  DSIS, PWC \\
\cline{5-6}
Su \textit{et al.}~\cite{su_perceptual_2019} & objects & static &   & \tabincell{l}{down-sampling, Gaussian noise,\\ V-PCC, S-PCC, L-PCC}  & \tabincell{c}{$20\times (3+9+9+12+4)=740$} &  & &  DSIS \\ 
\cline{5-6}
IRPC~\cite{javaheri_point_2019} & objects, humans & static & & PCL, G-PCC, V-PCC & $6\times 3\times 3=54$ & & & DSIS \\
\cline{5-6}

vsenseVVDB2~\cite{9123137} & humans & dynamic & & \tabincell{l}{Mesh: Draco+JPEG\\Point Clouds: G-PCC, V-PCC} & $8\times (6\times2+5)=136$ & & & ACR-HR \\
\cline{5-6}
Cao \textit{et al.}~\cite{9200318} & humans & dynamic & & \tabincell{l}{Mesh: TFAN + FFmpeg \\ Point Clouds: V-PCC + FFmpeg} & $4\times 1\times 5=20$ & & & ACR \\
\cline{5-6}\cline{7-7}
Stuart \textit{et al.}~\cite{9191308} & objects, humans & static & & G-PCC, V-PCC & $6\times (2+1)\times 5=90$ & 2D/3D monitor & & DSIS \\
\hline

Subramanyam \textit{et al.}~\cite{9089539} & humans & dynamic &\multirow{3}{*}{\checkmark} & the MPEG anchor, V-PCC & $8\times 2\times 4=64$ & \multirow{3}{*}{HMD} &\multirow{3}{*}{\checkmark}& ACR-HR \\
\cline{5-6}
PointXR~\cite{9123121} & objects & static &   & G-PCC  & $5\times (1+1)\times 4=40$ &  &  &  DSIS \\
\cline{5-6}
\textbf{Proposed SIAT-PCQD~\cite{siat-pcqd}} & objects, humans & static &   & V-PCC  & $20\times 1\times 17=340$ &  &  &  DSIS \\ \hline

\end{tabular}
}
\end{center}
\label{tbl:related work}
\end{table*}

Early works on point cloud subjective quality assessment came out in large numbers since 2014. In~\cite{zhang_subjective_2014}, distortions like down-sampling and noise generation were considered to study the relationship between 3D point cloud models and human visual perception. Later, some works
~\cite{javaheri_subjective_2017, alexiou_performance_2017, alexiou_impact_2018, alexious_point_2018, alexiou_point_2018-2} evaluated the quality of colorless point clouds and built the initial workflow of subjective point cloud evaluation. In~\cite{alexiou_towards_2017}~\cite{alexiou_impact_2018}, AR devices were first used in work about point clouds. 
However, only geometry of point clouds was evaluated in the subjective quality assessment. 
Also, point clouds were processed by Gaussian noise and octree-pruning degradation, and the latter promising point cloud codecs were failed to be considered. In the early stage, limited types of degradation were focused on, and some prevailing codecs were absent in these works.

Alexiou \textit{et al.}~\cite{alexiou_performance_2017} compared the Double Stimulus Impairment Scale (DSIS) and the Absolute Category Rating (ACR) methodology while evaluating point cloud geometry of Gaussian noise and octree-pruning degradation. They compared the visualization of raw point clouds and point clouds after surface reconstruction~\cite{alexious_point_2018}, and then obtained similar results using 2D and 3D monitors to display the projected contents of point clouds~\cite{alexiou_point_2018-2}. 
Javaheri \textit{et al.} performed a subjective and objective evaluation of point cloud denoising algorithms in~\cite{javaheri_subjective_2017}.
Different methodologies and ways of displaying were further studied in this period, but most of the subjective datasets still only used point cloud geometry as the evaluation contents.

Subsequently, colored point clouds under the primitive point cloud codec through octree-pruning and degradation like noise and down-sampling was further explored. 
Javaheri \textit{et al.}~\cite{javaheri_subjective_2017-1} performed subjective and objective PCQA by octree-based compression scheme, available in the Point Cloud Library (PCL), and graph-based compression scheme. They created a spiral virtual camera path moving around the point cloud sequences from a full view to a closer view, and the generated videos were evaluated by subjects. 
In~\cite{da_silva_cruz_point_2019}, point cloud evaluation experiments were conducted in three different laboratories, and it was found that removing points regularly was more acceptable for subjects. The quality scores, obtained by various point clouds with geometry and color information, showed a high correlation with objective metrics. 
Yang \textit{et al.}~\cite{9238424} proposed the SJTU-PCQA database with point clouds augmented with octree-based compression, color noise, geometry noise, and scaling, and they also developed an objective metric based on projection.

Later on, the advanced point cloud codecs developed by MPEG were introduced in subjective PCQA studies.
In~\cite{zerman_subjective_2019}, Zerman \textit{et al.} first considered V-PCC for colored point clouds and rendered point clouds by Unity in the way of no interaction. They found that texture distortion is more critical than geometric distortion in the human figure database they created. They also found that the count of points severely affects geometric quality metrics rather than perceptual quality. Su \textit{et al.}~\cite{su_perceptual_2019} built a point cloud database of diverse contents and applied down-sampling, Gaussian noise, and three state-of-the-art PCC algorithms to create distorted point clouds. They first explicitly defined types of distortions for point clouds: geometry distortions include hollow, geometry noise, hole, shape distortion, collapse, and gap and blur; texture distortions include texture noise, blocking, blur, and color bleeding. 
Javaheri \textit{et al.}~\cite{javaheri_point_2019} created a subjective database named IRPC and studied the impacts of different coding and rendering solutions on the perceptual quality. 
For the comparison of dynamic point clouds and meshes, the perceptual quality of compressed 3D sequences was explored~\cite{9123137}~\cite{9200318}, and Cao \textit{et al.}~\cite{9200318} first study the impact of observation distance on perceptual quality. 
Recently, Perry~\textit{et al.}~\cite{9191308} confirmed the superior compression performance of the MPEG V-PCC compared to MPEG G-PCC in static contents. 
Thus, researchers considered prevailing point cloud codecs like V-PCC and G-PCC on diverse colored point clouds. However, point clouds were rendered as video sequences by individual tracks around the point cloud centers and then displayed on a planar screen in passive interaction. 

As for interaction, a quality evaluation methodology for colored and voxelized point clouds was proposed in~\cite{torlig_novel_2018}. In their experiment, subjects visualized the contents through renderer and interacted with point clouds by zooming, rotating, and translation using the mouse.
Alexiou \textit{et al.}~\cite{alexiou_comprehensive_2019} focused on the evaluation of test conditions defined by MPEG for core experiments and conducted two additional rate allocation experiments for geometry and color encoding modules. A new software, which supports interaction and would be used in the web applications for point cloud rendering, was developed. In a word, these works enabled interaction by allowing subjects to operate on point clouds displayed in a 2D monitor, namely the desktop condition. 
With the rapid development of computer graphics and progressive technologies, 3D data can be fully exhibited in VR applications. User quality evaluation of dynamic point clouds in VR  was performed in the works~\cite{9089539}~\cite{9123121}. Subramanyam \textit{et al.}~\cite{9089539} first compared two VR viewing conditions enabling 3DoF and 6DoF by assessing the quality of dynamic digital human point clouds, and the latter work~\cite{9123121} developed the PointXR toolbox that can host experiments under variants of protocols in the VR environment.

In our work, a subjective quality evaluation experiment was conducted with positive interaction in the 6DoF VR environment. 
We aim to explore the impact of degradation of point clouds' geometry and texture attributes for visual quality in future VR applications. 
Thus, seventeen distorted rates for compressed degradation are set in our subjective experiment where the number of rate levels is far more than other works. A summary of the available subjective PCQA databases is shown in Table~\ref{tbl:related work}. 

\subsection{Objective Point Cloud Quality Assessment}
Objective quality assessment of point clouds aims to create an accurate mathematical model to predict the quality of point clouds. 
According to the work~\cite{torlig_novel_2018}, the state-of-the-art objective evaluation metrics of point clouds can be classified into two categories: point-based metrics and projection-based metrics. 

Point-based metrics are mainly the point-to-point error (D1)~\cite{N17995}, the point-to-plane error (D2)~\cite{tian_geometric_2017} and the plane-to-plane error~\cite{alexiou_point_2018} for geometric errors. D1 measures the Euclidean distances between corresponding point pairs, indicating how far the distorted point cloud points moved away from their original positions. Considering local plane properties, D2~\cite{tian_geometric_2017} computes the projected errors along the normal direction on the associated point, imposing a larger penalty on points far from the perceived local plane surface. 
The plane-to-plane metric~\cite{alexiou_point_2018} focuses on the angular similarity between tangent planes of associated points, and tangent planes indicate the linear approximation of the surface. All of them are full-reference metrics based on geometric errors, regardless of the human vision.

To further improve the prediction accuracy, point-based metrics have been explored by extracting features in geometry and texture. 
Alexiou \textit{et al.}~\cite{alexiou_point_2018} introduced an objective quality metric based on the angular similarity between tangent planes. 
Meynet \textit{et al.}~\cite{meynet_pc-msdm_2019} extended the Mesh Structural Distortion Measure (MSDM) metric designed for 3D meshes to point clouds as PC-MSDM. 
Javaheri \textit{et al.} successively proposed novel geometry quality metrics for point clouds based on the popular geometry quality metric PSNR~\cite{9191233}, a generalization of the Hausdorff distance~\cite{9123087}, and the Mahalanobis distance to measure the correspondence between a point and a distribution~\cite{9143408}, respectively. 
After exploring the geometry quality of point clouds, researchers focused on extracting color features and combining the geometry and the color domain. 
Rafael \textit{et al.} computed the direct distance between points~\cite{9190956}. They adapted the Local Binary Pattern (LBP) descriptor to processing local regions, and the histograms of the LBP outputs were compared to obtain the final score~\cite{9123076}. 
Meynet \textit{et al.} linearly combined geometry-based features, i.e., curvature, and color-based features, i.e., lightness, chroma, and hue, and proposed the Point Cloud Quality Metric (PCQM)~\cite{9123147}. 
Color histograms and color correlograms were utilized and combined with geometry metrics to provide a global quality score~\cite{9123089}. 
Yang \textit{et al.} resampled the reference point cloud to extract key points, constructed a local graph centered at key points, and aggregated color gradient features to form the method named GraphSIM~\cite{yang2020inferring}. 
Alexiou \textit{et al.}~\cite{9106005} focused on structual similarity and local distributions of point cloud attributes reflecting topology and color. 
Viola \textit{et al.}~\cite{9198142} proposed a reduced reference metric by extracting statistical features in geometry, color, and normal vector domain. 
However, the operation of point clouds as nonstructural data, like computing curvature and resampling, is time-consuming and computationally expensive.

Projection-based metrics project both reference and test point clouds onto six planes of their bounding boxes~\cite{torlig_novel_2018} or more planes, then compute the average scores of the structural data, i.e., projected images, by the state-of-the-art image quality metrics.
In~\cite{alexiou_exploiting_2019}, projection-based objective quality assessment was extended by assigning weights to perspectives based on user interactivity data. The authors identified that there is not much difference when using additional views. Yang \textit{et al.}~\cite{9238424} proposed a projection-based method via perspective projection onto six planar and extracted global and local features of depth and color images obtained by projection. However, occlusion and dislocation are inevitable for projecting point clouds onto large planes. 
Thus, we build a subjective PCQA database in the immersive 6DoF VR environment and propose two projection-based objective evaluation methods.

\section{Subjective Experiment for PCQA}\label{section3}
In this section, we describe the details of our point clouds subjective quality assessment experiment with dataset preparation, processing, equipment and test environment, and evaluation methodology. 

\begin{figure*}
\centerline{\includegraphics[width=17.6cm]{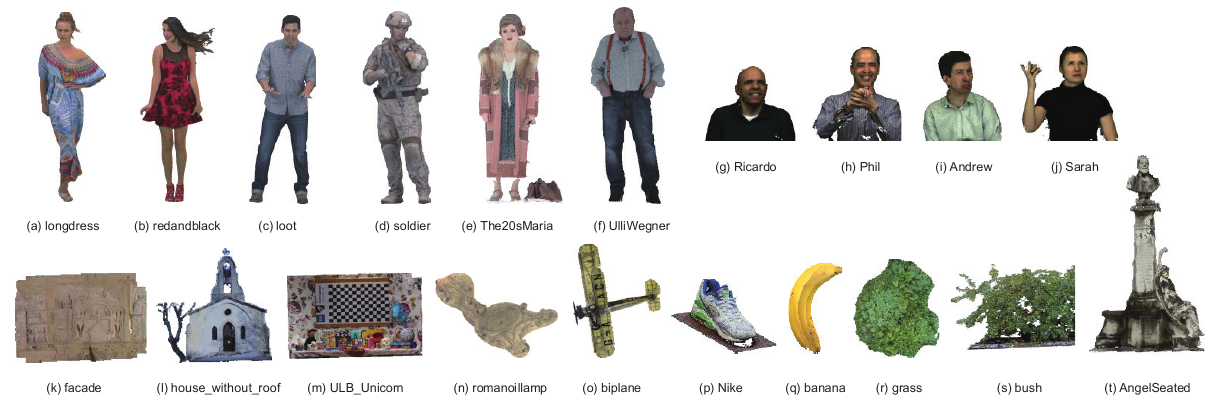}}
\caption{The thumbnail images of the sequences used in our experiment.}
\label{fig:sequences}
\end{figure*}

\begin{table*}
\caption{Summary of Pre-processed Test Sequences.}
\begin{center}
\resizebox{\textwidth}{!}{
\begin{threeparttable}
\begin{tabular}{l|c|c|c|c|c|c|c}
\hline
Sequence & Category & Source & \tabincell{l}{Pre- \\ processing} & \#Points & \tabincell{l}{Geometry \\ Precision}& Bounding Box & \tabincell{l}{(geometry QP, \\ texture QP)} \\
\hline
Redandblack~\cite{seq:8i} & Full body figures & MPEG\footnotemark[1]/JPEG\footnotemark[2] & No &$729,133$ & 10 bits & $(393, 977, 232)$ & \multirow{20}{*}{\tabincell{l}{(20,27), (20,37),\\ (20,47), (28,27),\\ (28,37), (28,47),\\ (36,27), (36,37),\\ (36,47), (24,32),\\ (32,42), \textbf{(0,0)},\\ (20,0), (28,0),\\ (36,0), (0,27),\\ (0,37), (0,47)}}\\
Longdress~\cite{seq:8i} & Full body figures & MPEG\footnotemark[1]/JPEG\footnotemark[2] & No & $765,821$ & 10 bits & $(356, 1003, 296)$\\
Loot~\cite{seq:8i} & Full body figures & MPEG\footnotemark[1]/JPEG\footnotemark[2] & No & $784,142$ & 10 bits & $(352, 992, 354)$ \\
Soldier~\cite{seq:8i} & Full body figures & MPEG\footnotemark[1]/JPEG\footnotemark[2] & No & $1,059,810$ & 10 bits & $(360, 1016, 405)$\\
The20sMaria~\cite{seq:maria} & Full body figures & MPEG\footnotemark[1] & Yes & $950,423$ & 10 bits & $(405, 908, 324)$ \\
UlliWegner~\cite{seq:wegner} & Full body figures & MPEG\footnotemark[1] & Yes & $598,448$  & 10 bits & $(376, 997, 258)$\\
Ricardo~\cite{seq:microsoft} & Upper body figures & JPEG\footnotemark[2] &  No &$960,703$ & 10 bits & $(446, 364, 178)$ \\
Phil~\cite{seq:microsoft} &  Upper body figures  & JPEG\footnotemark[2] & No & $1,660,959$ & 10 bits & $(441, 464, 394)$ \\
Andrew~\cite{seq:microsoft} &  Upper body figures  & JPEG\footnotemark[2] & No & $1,276,312$ & 10 bits & $(392, 444, 297)$ \\
Sarah~\cite{seq:microsoft} &  Upper body figures  & JPEG\footnotemark[2] & No & $1,355,867$ & 10 bits & $(486, 467, 348)$\\
\cline{1-7}
Facade & Inanimate objects  & MPEG\footnotemark[1] & Yes & $292,169$ & 10 bits & $(555, 375, 75)$ \\
House\_without\_roof & Inanimate objects  & MPEG\footnotemark[1] & Yes & $581,213$ & 10 bits & $(488, 481, 455)$\\
ULB\_Unicorn & Inanimate objects  & MPEG\footnotemark[1] & Yes & $1,086,944$ & 10 bits & $(571, 361, 303)$\\
Romanoillamp~\cite{seq:roman} & Inanimate objects & JPEG\footnotemark[2] & Yes & $343,186$ & 10 bits & $(517, 355, 352)$ \\
Biplane~\cite{seq:biplane} & Inanimate objects & JPEG\footnotemark[2] & Yes & $400,972$ & 10 bits & $(439, 569, 410)$\\
Nike & Inanimate objects & Sketchfab\footnotemark[3] & Yes & $186,960$ & 10 bits & $(303, 213, 303)$\\
Banana & Inanimate objects & Sketchfab\footnotemark[3] & Yes & $145,243$ & 10 bits & $(201, 337, 102)$\\
Grass & Inanimate objects & Sketchfab\footnotemark[3] & Yes & $724,725$ & 10 bits & $(494, 159, 434)$\\
Bush & Inanimate objects & Sketchfab\footnotemark[3] & Yes & $1,211,816$ & 10 bits & $(587, 400, 435)$ \\
AngelSeated & Inanimate objects & Sketchfab\footnotemark[3] & Yes & $770,184$ & 10 bits & $(543, 942, 305)$ \\
\hline
\multicolumn{8}{l}{Bold number denotes lossless compression.}
\end{tabular}
   \begin{tablenotes}
     \item[1]{https://mpeg.chiariglione.org/tags/point-cloud}
     \item[2]{https://jpeg.org/plenodb/}
     \item[3]{https://sketchfab.com/} 
   \end{tablenotes}
\end{threeparttable}
}
\end{center}

\label{tbl:sequences}
\end{table*}

\subsection{Dataset Preparation}

To better explore how people perceive point clouds, we chose contents like human figures and inanimate objects. As shown in Fig.~\ref{fig:sequences} and Table~\ref{tbl:sequences}, 20 static sequences were selected in our test. The human category consists of six full-body figures and four upper-body figures, while another category includes ten different inanimate objects. 


\begin{figure}
\centerline{\includegraphics[width=0.8\linewidth]{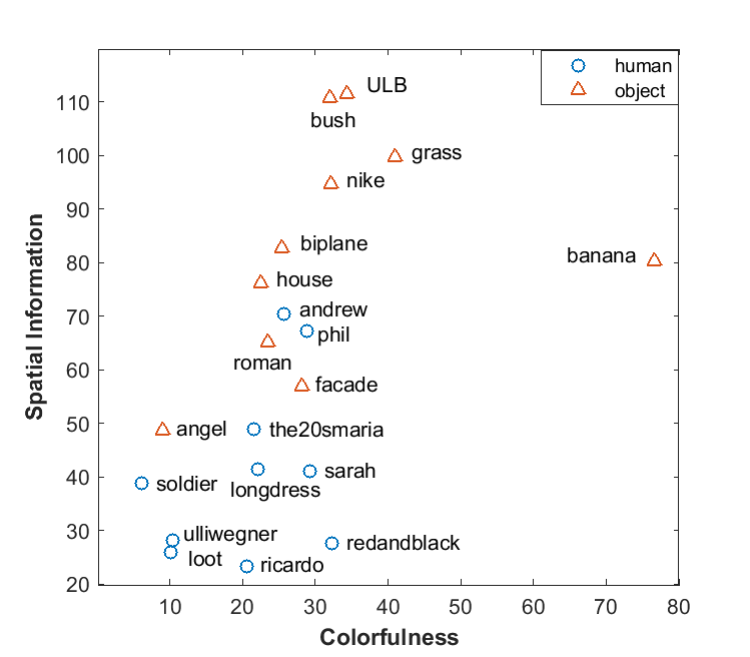}}
\caption{Distribution of spatial information and colorfulness of the 20 source sequences in the dataset.}
\label{fig:SI_CF}
\end{figure}

To show the diversity of point clouds, we considered the characteristics, i.e., Spatial Information (SI)~\cite{itu910} and ColorFulness (CF)~\cite{6280595}.
We projected the source point cloud into six views of its bounding box to apply SI and CF. Similar to video contents in~\cite{itu910}, we obtained the maximum value among six views as the final SI for a sequence. Fig.~\ref{fig:SI_CF} shows the distribution of 20 sequences along with the horizontal (CF) and vertical (SI) axes. The dispersed state in CF/SI shows the diversity of our contents in space/color domain. In particular, the luminance of the sequence \textit{Banana} is generally higher than others, making a CF measurement difference.

\subsection{Processing}

\begin{figure}
\includegraphics{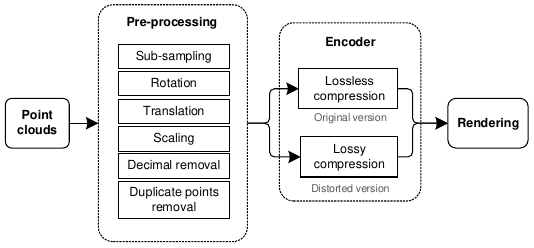}
\caption{The workflow before conducting the subjective experiment, including preprocessing, encoding, and rendering.}
\label{fig:process}
\end{figure}

Before compression, it requires the preprocessing procedure for point cloud sequences to minimize the impact of some additional influencing factors. Fig.~\ref{fig:process} shows the whole workflow before conducting the experiment, which mainly includes preprocessing, encoding, and rendering.

\begin{itemize}[leftmargin = 0pt, itemindent = 2em]
\item \textbf{Preprocessing:} The sequences, as mentioned above, are selected from different repositories, which means their sizes, positions, and orientations vary. However, we desire that point clouds are exhibited in life-size rendering to achieve realistic tele-immersive scenarios. So we normalize sequences to remain point clouds within a similar bounding box $(600, 1000, 400)$ in the preprocessing stage to deal with this issue. The source models have been processed with sub-sampling, rotation, translation, and scaling, except four sequences \textit{Longdress}, \textit{Redandblack}, \textit{Loot}, and \textit{Soldier} from the 8i Voxelized Full Bodies Database. 
Additionally, the point cloud encoder V-PCC fails to deal with decimals, so that the positions of points were through round operation and then the duplicate points were removed. 
In particular, it is unnecessary to have integer conversion for four upper body figure sequences from the Microsoft database, so we just adjusted their positions and orientations in rendering software.

\item \textbf{Encoding:} Distorted versions were generated using the state-of-the-art MPEG PCC reference software Test Model Category 2 version 7.0 (TMC2v7.0). The V-PCC method takes advantage of an advanced 2D video codec after projecting point clouds into frames. First, a point cloud is split into patches by clustering normal vectors. The obtained patches would be packed into images, and the gaps between patches would be padded to reduce pixel residuals. Then projected images of sequences are compressed utilizing the HEVC reference software HM16.18. More information about the framework of V-PCC can be referred to~\cite{schwarz_emerging_2019}.
\\
\hspace*{1em}
QP determines the step size for transformed coefficients in codecs. In V-PCC, a pair of parameters, namely geometry QP and texture QP, regulate how much detail is saved in the geometry and texture attributes of point clouds. As geometry QP is increased, points deviate from their original positions. As texture QP is increased, some color details are aggregated. 
Similar to the Common Test Conditions (CTC) document from the MPEG PCC~\cite{N17995}, the gaps of geometry and texture QPs were set as 4 and 5, ranging from 20 to 32 and from 27 to 42. 
As shown in Table~\ref{tbl:sequences}, geometry QP ranks the first in each pair, while texture QP ranks the second in that pair. And we chose a losslessly compressed version as our reference contents. 

\item \textbf{Rendering:} Point clouds are appropriate to represent the complete view of objects and scenes in immersive applications with 6DoF. 
Thus, we developed an actively interactive VR experiment software for subjects to observe point cloud models in the 6DoF environment. Observers are permitted to explore freely in a room and observe the point clouds from any angle without occlusion. In this condition, the views displayed within the HMD are consistent with observers' body and head movements, resulting in an immersive feeling of perception.
\\
\hspace*{1em}
Our experiment software was developed in Unity (version 5.6.5f1), exploiting the SteamVR plugin (version 1.2.3) to connect VR headsets. Point Cloud Viewer and Tools (version 2.20) helped us import and view the point cloud data inside Unity. 
After preprocessing, point clouds are rescaled to a similar size to exhibit realistic tele-immersive scenarios. 
Besides, geometric coding distortions can be masked by surface reconstruction~\cite{javaheri_point_2019}. Thus, raw point clouds are presented using the default point size. 
Notably, a large size of point cloud files might take up too much memory and cause a system hang. So we first packed all the resources related to rendering point clouds like prefabs and meshes and dynamically asynchronously loaded the asset bundles to improve software stability. 
\end{itemize}

\subsection{Equipment and Environment}
\begin{figure*}
\centering
\subfigure[]{
    \begin{minipage}[t]{0.14\linewidth}
    \centering
    \includegraphics[width=1\linewidth]{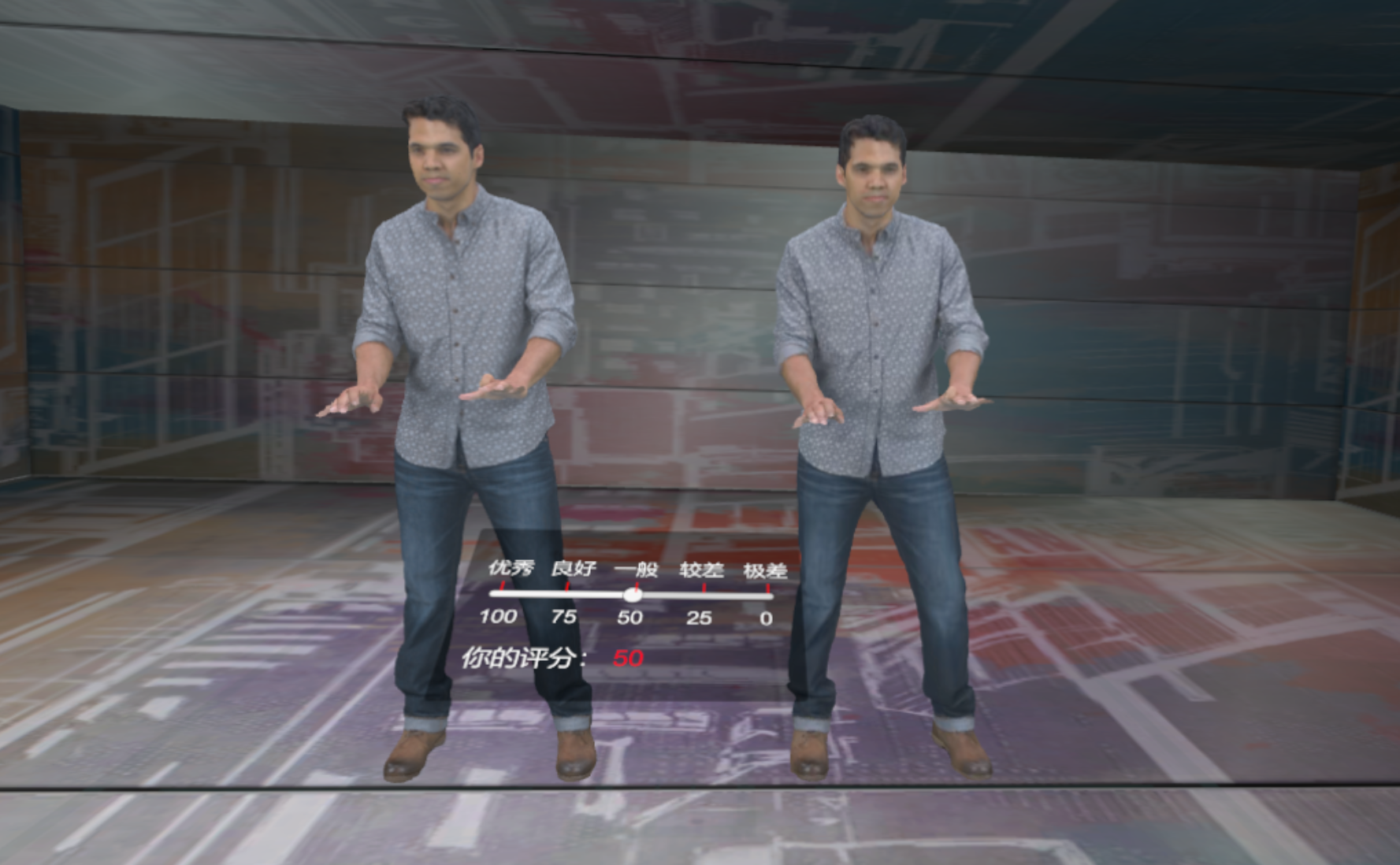}
    \end{minipage}
}
\subfigure[]{
    \begin{minipage}[t]{0.14\linewidth}
    \centering
    \includegraphics[width=1\linewidth]{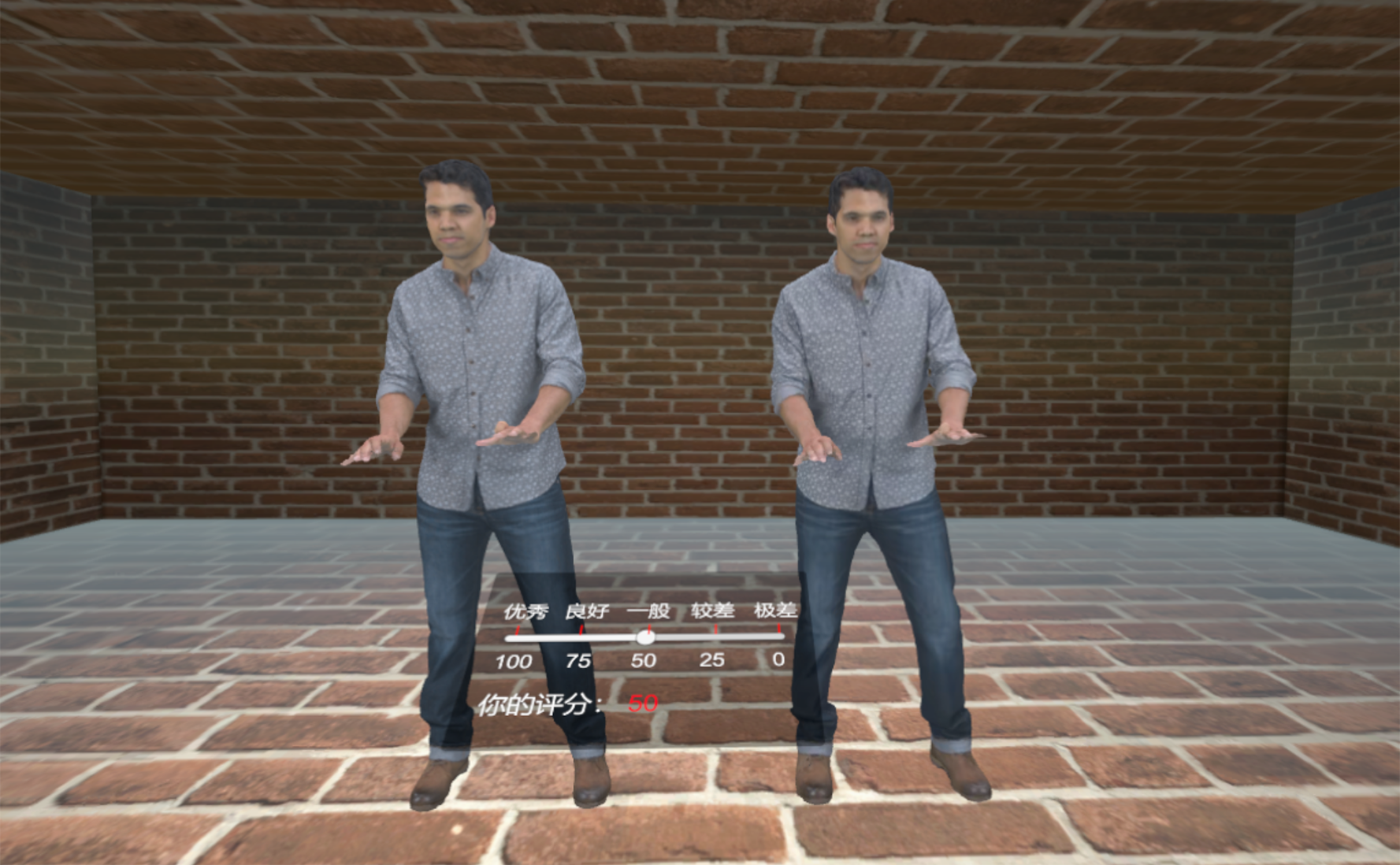}
    \end{minipage}
}
\subfigure[]{
    \begin{minipage}[t]{0.14\linewidth}
    \centering
    \includegraphics[width=1\linewidth]{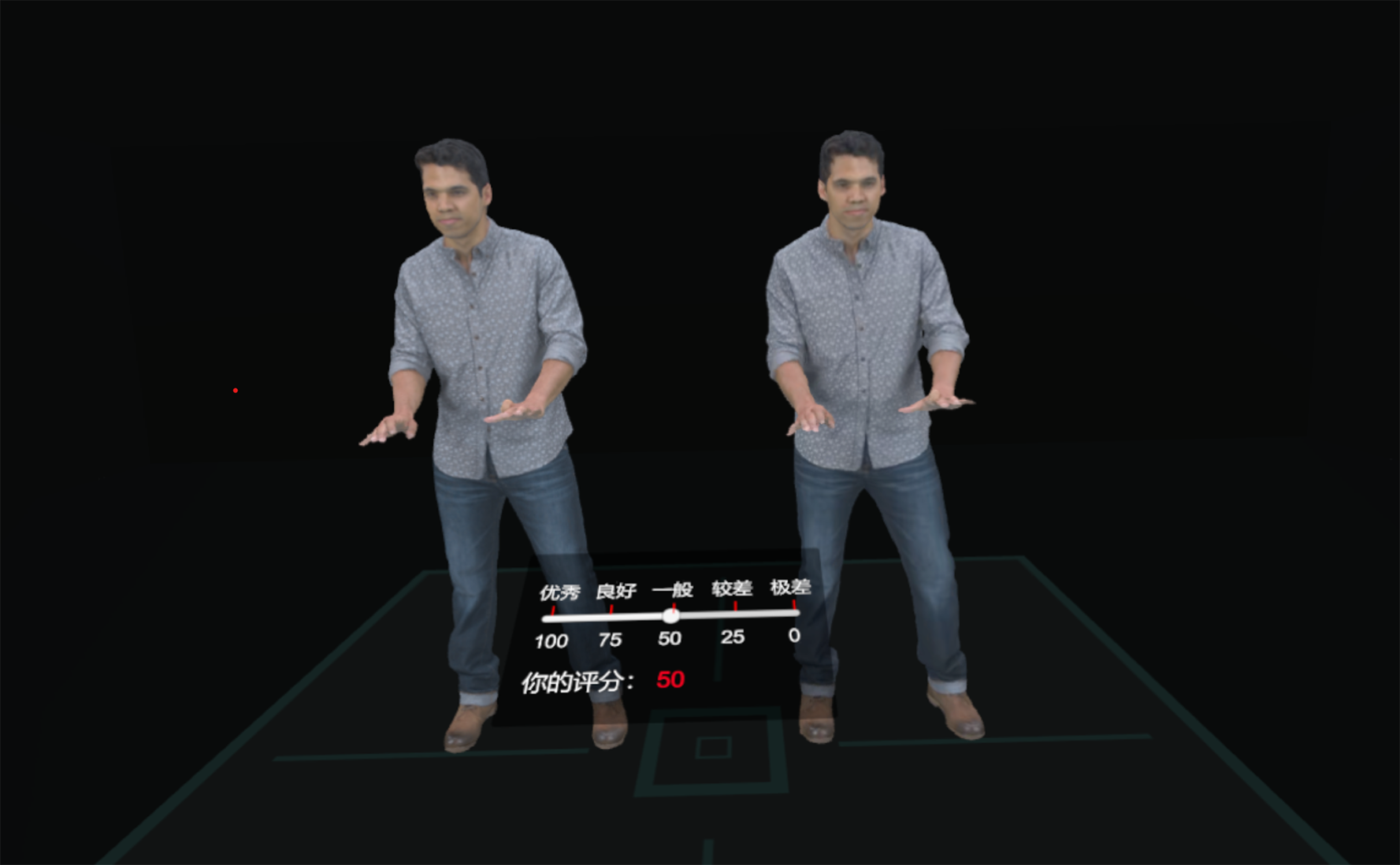}
    \end{minipage}
}
\subfigure[]{
    \begin{minipage}[t]{0.14\linewidth}
    \centering 
    \includegraphics[width=1\linewidth]{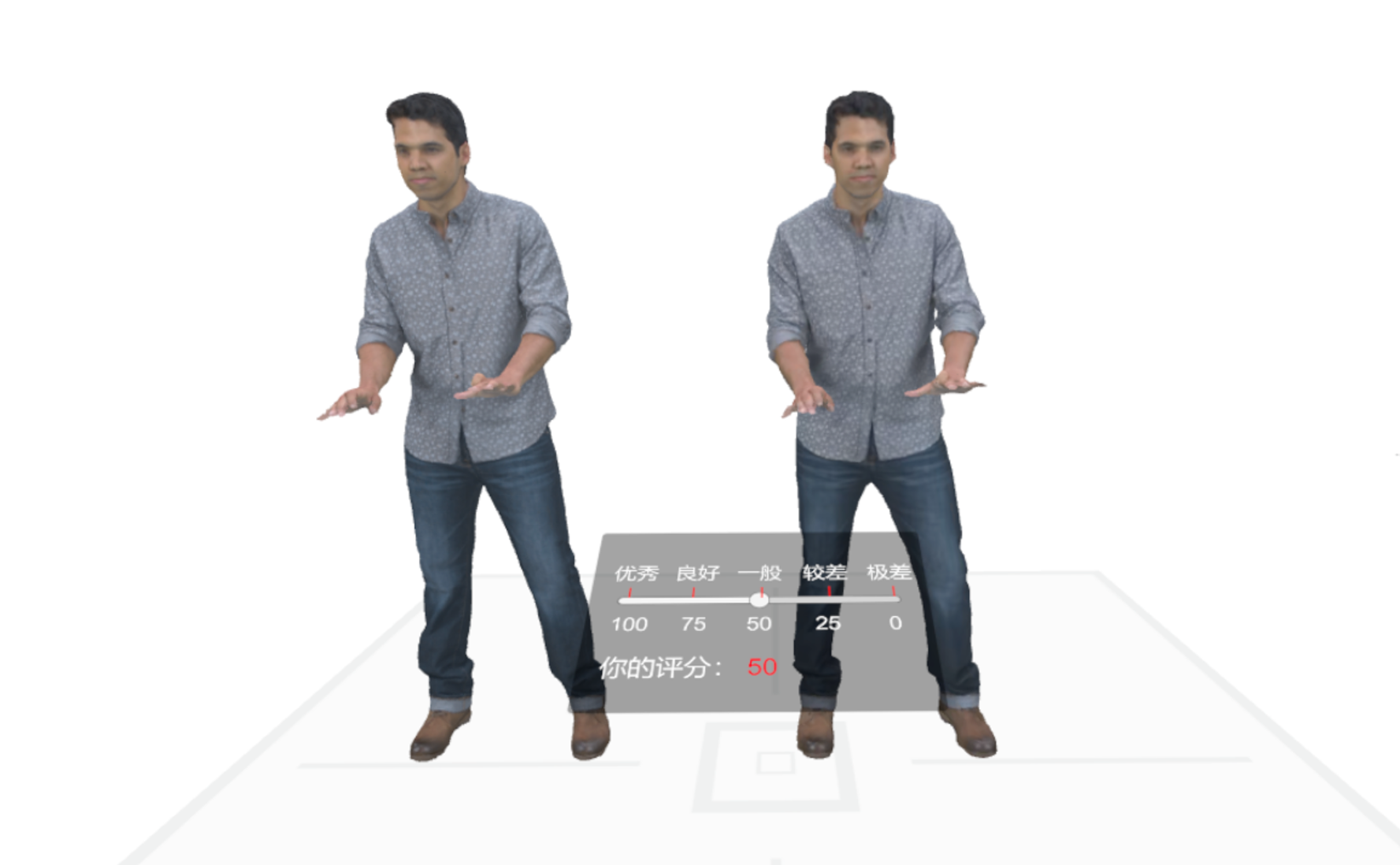}
    \end{minipage}
}
\subfigure[]{
    \begin{minipage}[t]{0.14\linewidth}
    \centering 
    \includegraphics[width=1\linewidth]{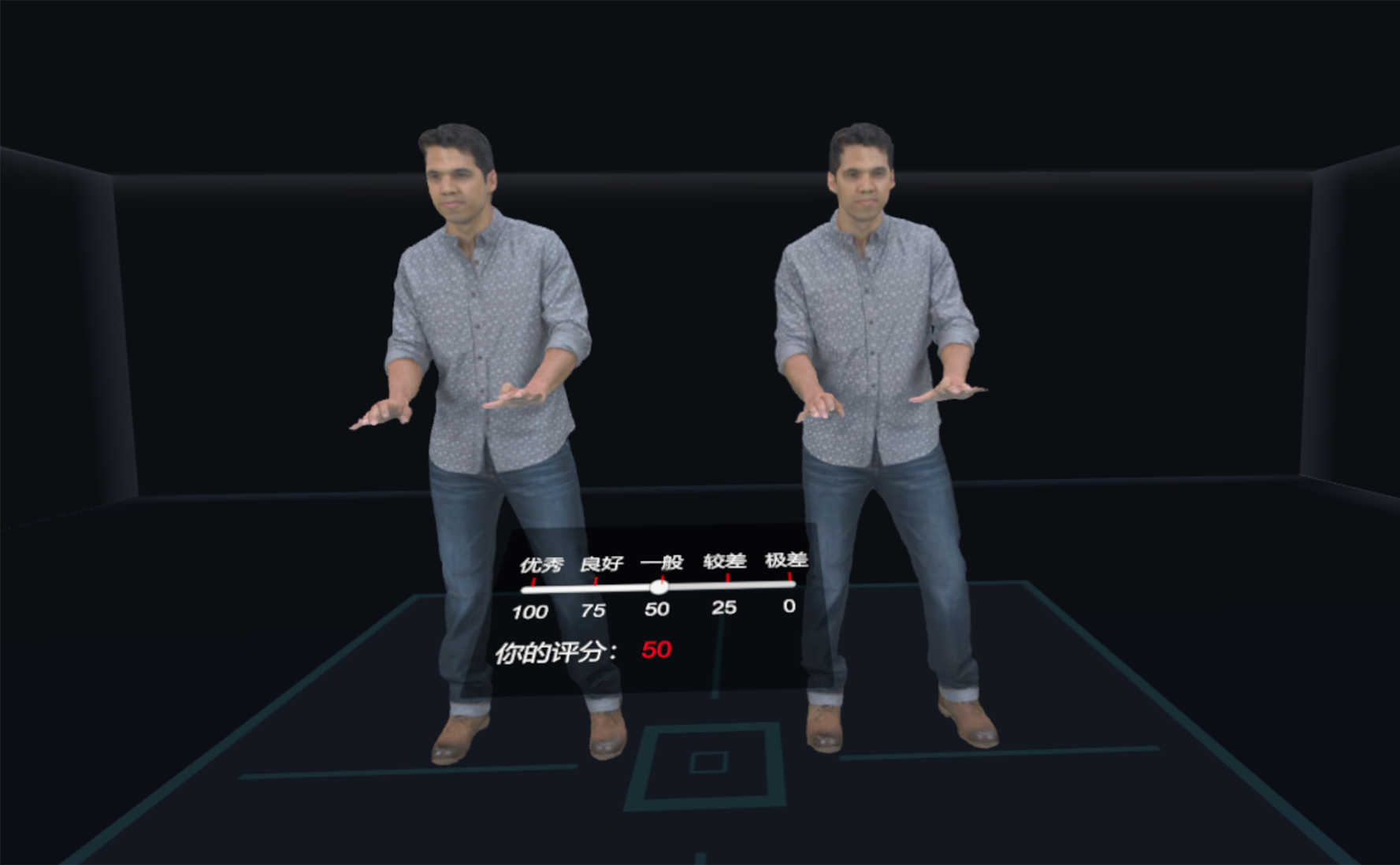}
    \end{minipage}
}
\\
\subfigure[]{
    \begin{minipage}[t]{0.14\linewidth}
    \centering 
    \includegraphics[width=1\linewidth]{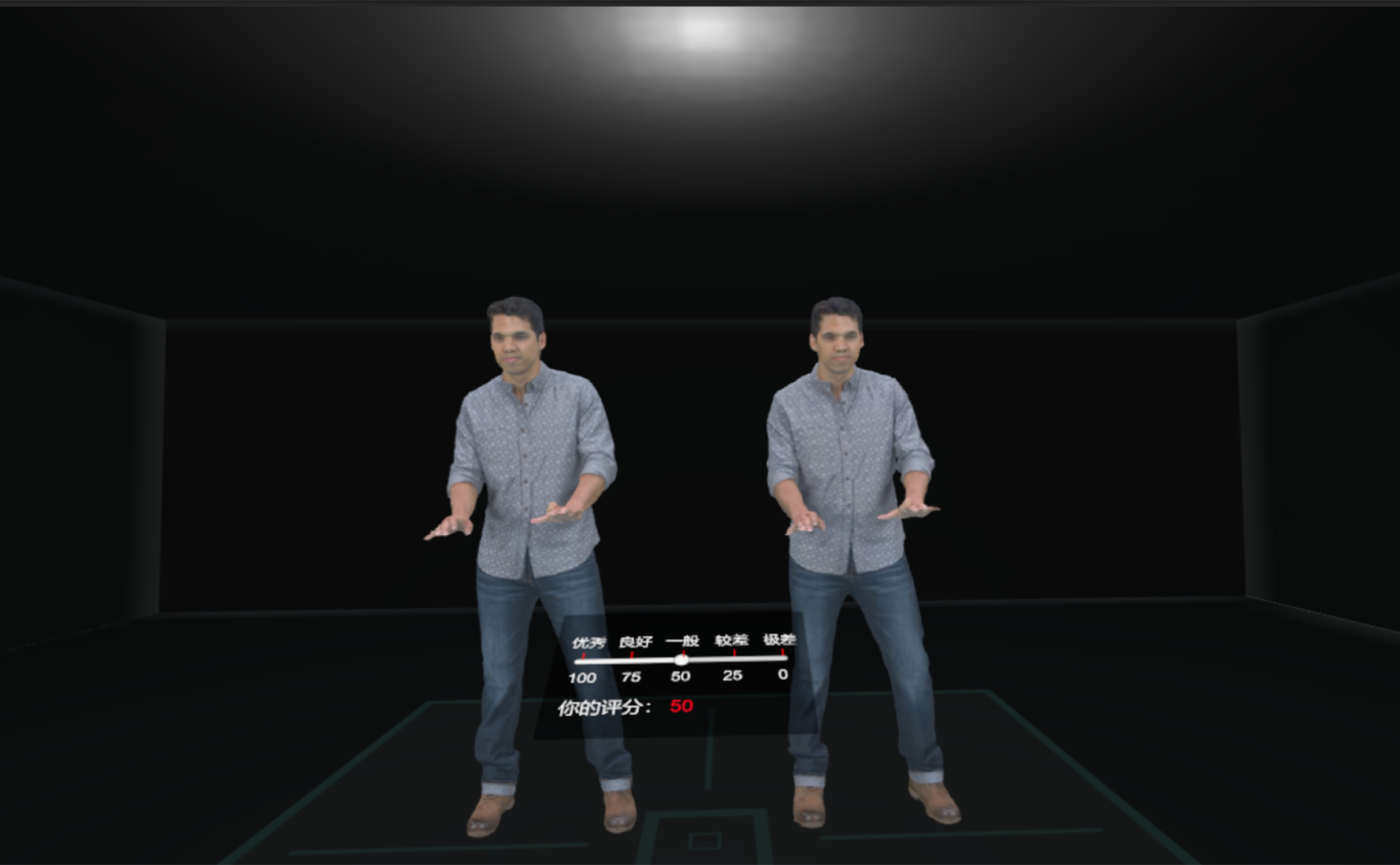}
    \end{minipage}
}
\subfigure[]{
    \begin{minipage}[t]{0.14\linewidth}
    \centering 
    \includegraphics[width=1\linewidth]{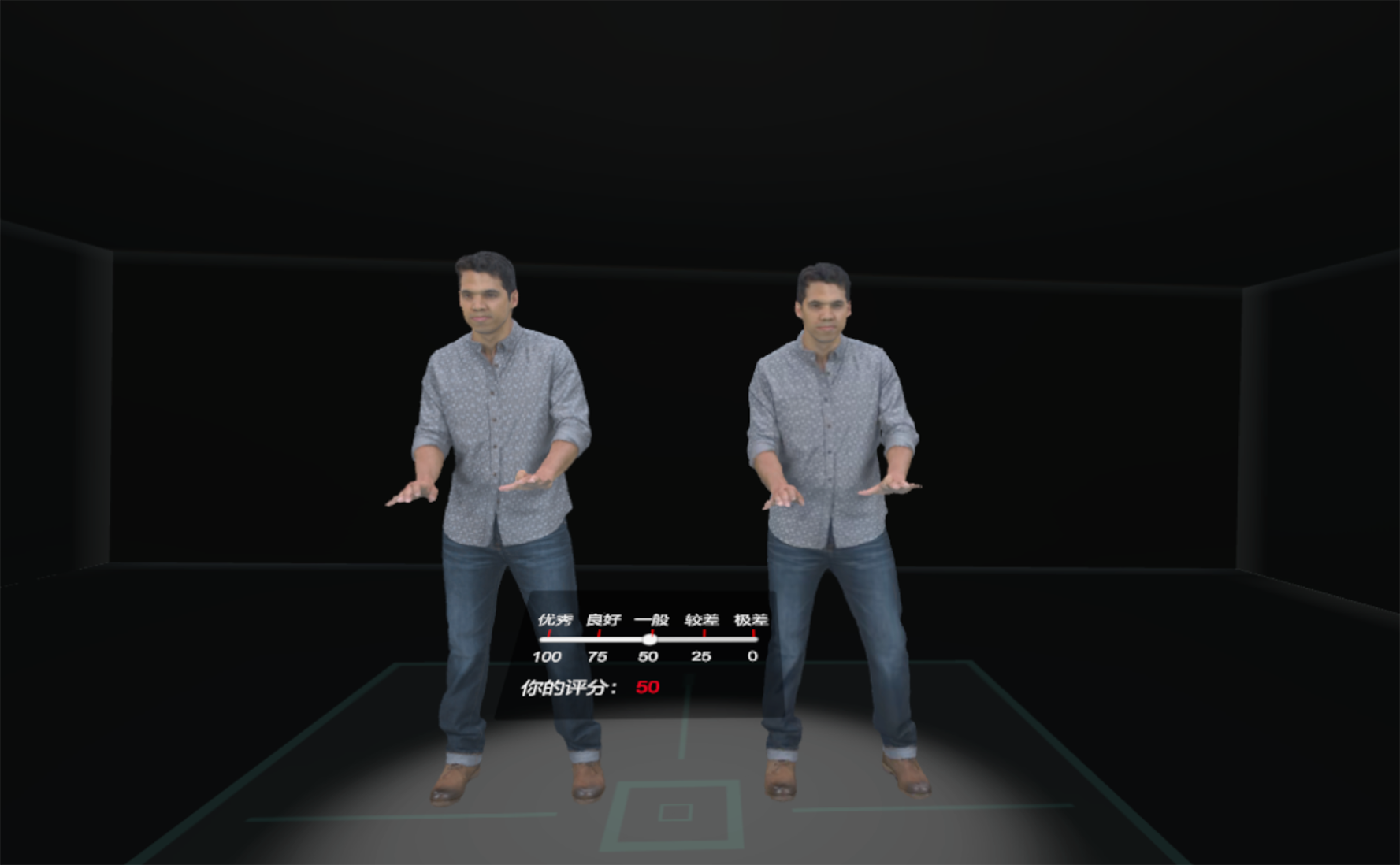}
    \end{minipage}
}
\subfigure[]{
    \begin{minipage}[t]{0.14\linewidth}
    \centering 
    \includegraphics[width=1\linewidth]{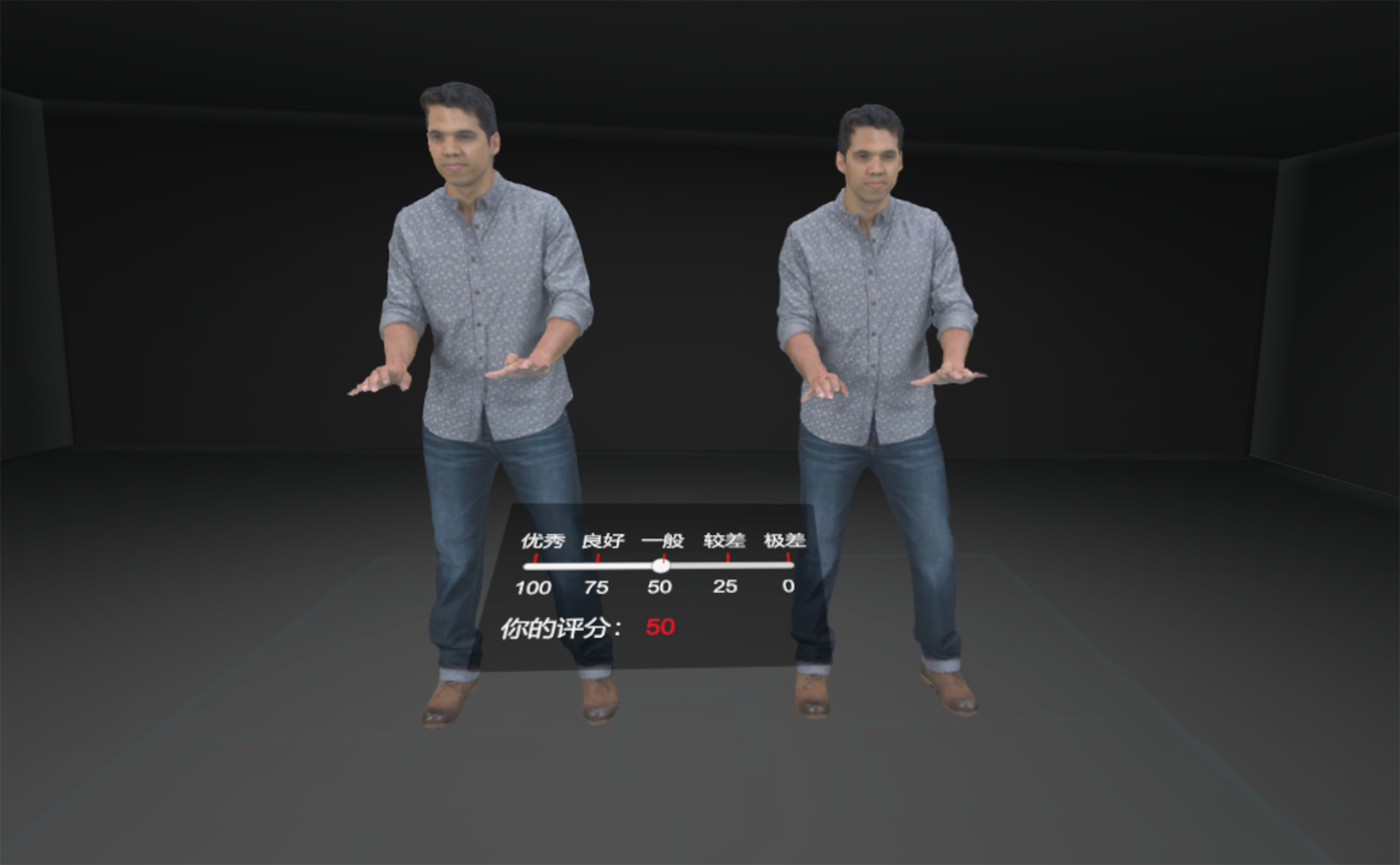}
    \end{minipage}
}
\subfigure[]{
    \begin{minipage}[t]{0.14\linewidth}
    \centering 
    \includegraphics[width=1\linewidth]{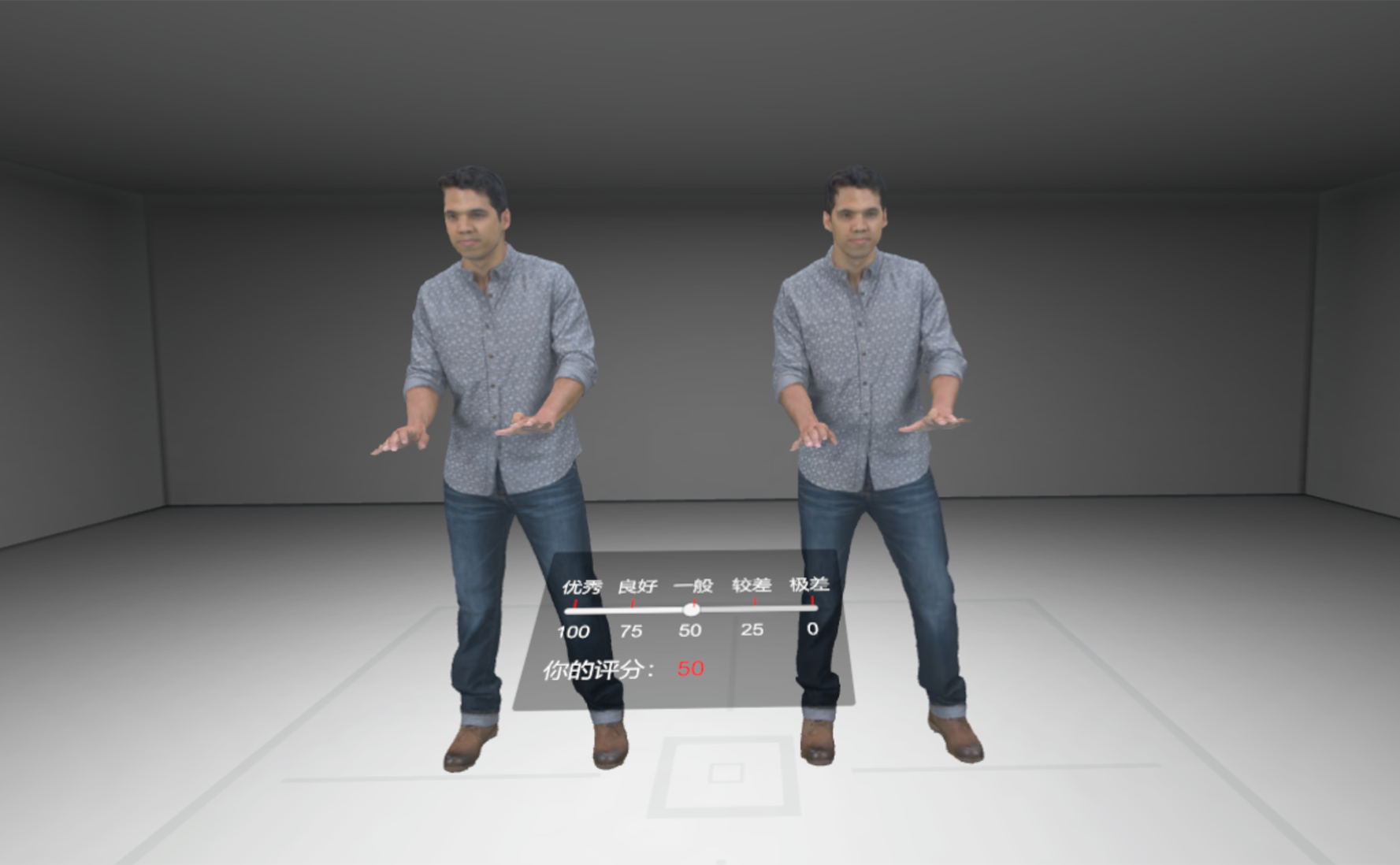}
    \end{minipage}
}
\subfigure[]{
    \begin{minipage}[t]{0.14\linewidth}
    \centering 
    \includegraphics[width=1\linewidth]{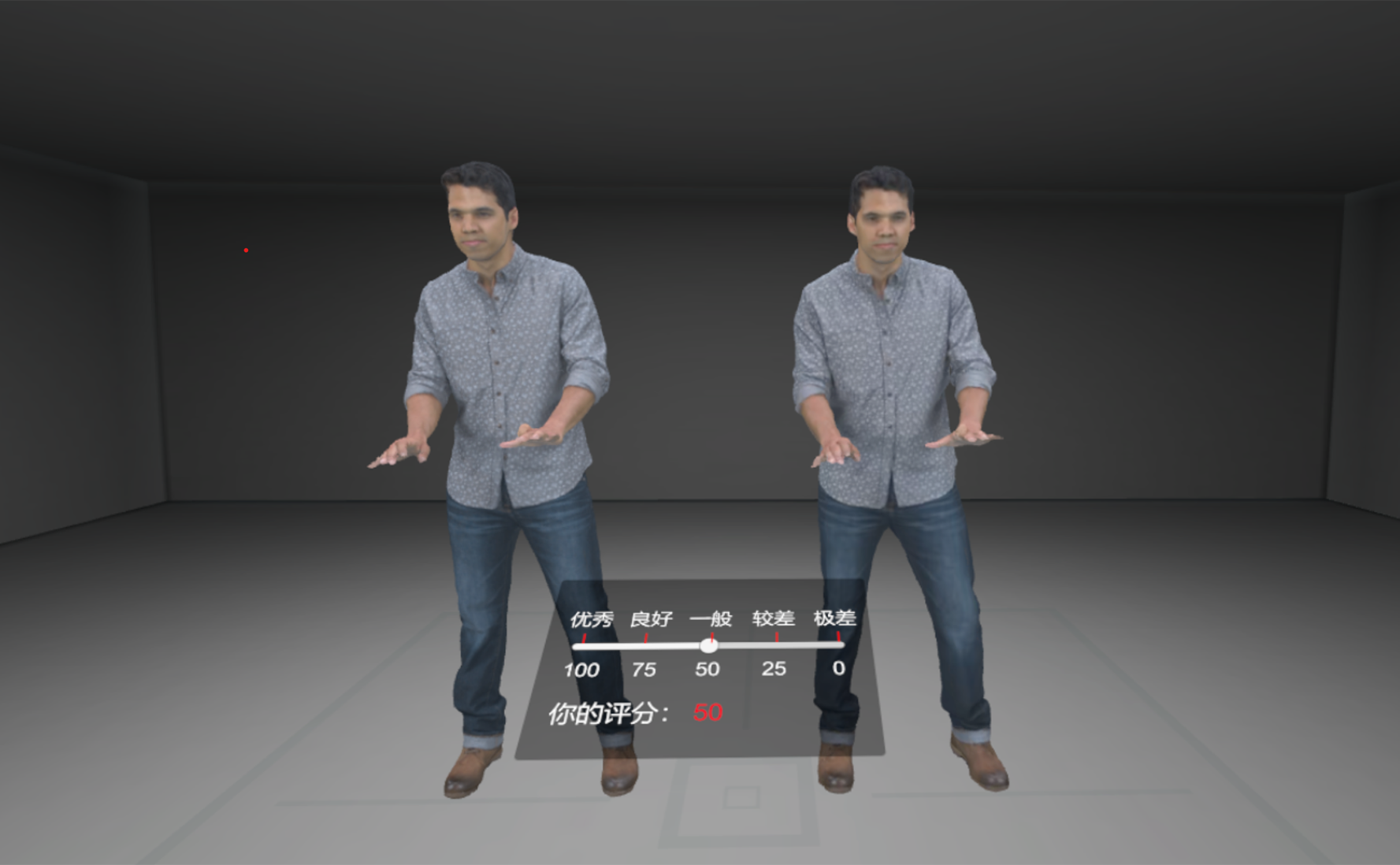}
    \end{minipage}
}

\caption{Scene settings in the virtual environment. (a) wallpaper background, (b) brick background, (c) black background, (d) white background, (e) directional light, (f) point light, (g) spot light, (h) light intensity = 1, (i) light intensity = 2, (j) the adopted setting.
}  
\label{fig:environment}
\end{figure*}

HTC Vive devices with a HMD and two hand controllers were used for every subject to interact in our test. The headset features a resolution of $1080\times1200$ pixels per eye, namely $2160\times1200$ pixels, and a 110-degree field of view. 
As for the virtual environment setting, backgrounds with complex textures or contrasting colors and settings with high or low light intensity would enhance the contrast between point clouds and the environment, as shown in Fig.~\ref{fig:environment}. Thus, we preferred a comfortable environment with mild contrast as the scene of our experiment. 
Following the recommendations from Recommendation ITU-R BT.500-13~\cite{itu500}, a room with gray walls was created as the virtual environment to conduct the experiment. Meanwhile, the scene was lighted by a virtual lamp on the ceiling centralized above the models. 
The lamp is set as an area light with intensity values of 2 in Unity to simulate ordinary lighting in the room. 

\begin{figure}
\centering

\subfigure[]{
    \begin{minipage}[t]{0.295\linewidth}
    \centering
    \includegraphics[width=1\linewidth]{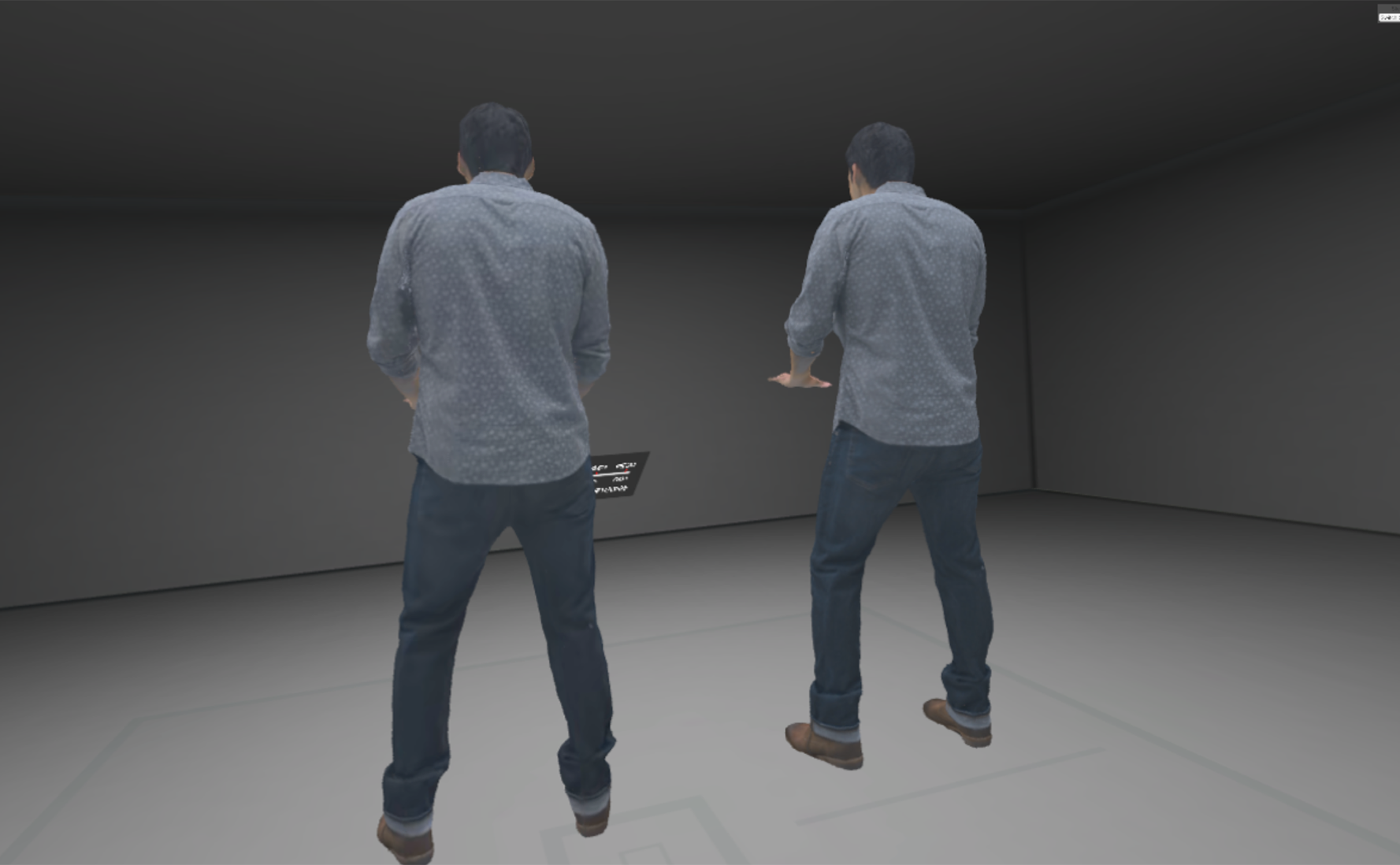}
    \end{minipage}
}
\subfigure[]{
    \begin{minipage}[t]{0.295\linewidth}
    \centering
    \includegraphics[width=1\linewidth]{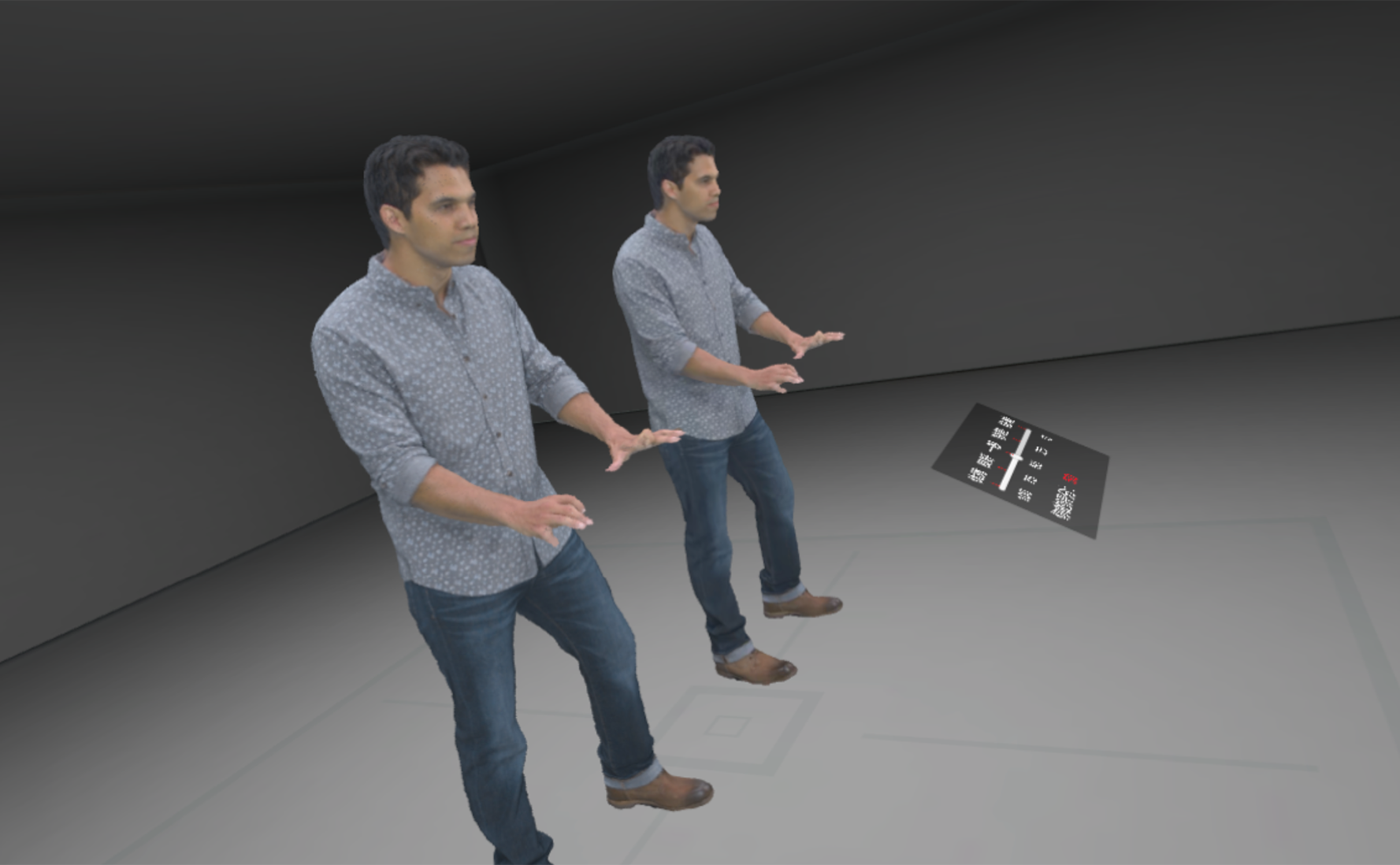}
    \end{minipage}
}
\subfigure[]{
    \begin{minipage}[t]{0.295\linewidth}
    \centering 
    \includegraphics[width=1\linewidth]{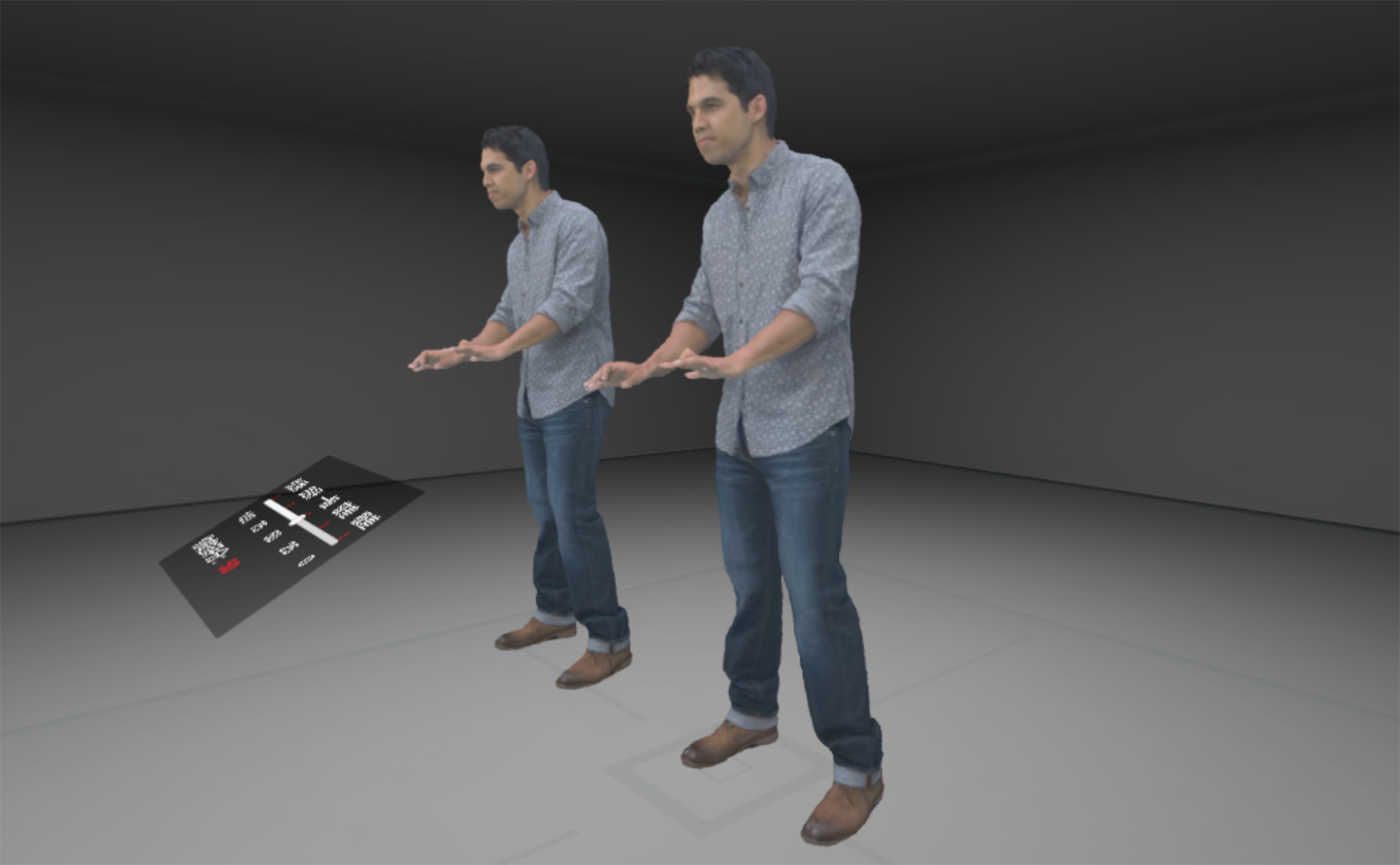}
    \end{minipage}
}
\caption{Views of the experiment system in HMD from some angles. (a) back, (b) left, (c) right.
}  
\label{fig:screenshot}
\end{figure}

For 2D images and videos, the best way of presentation is displaying on the planar screen. For volumetric media like meshes and point clouds, it is more natural to exhibit the 3D models in virtual 3D worlds. In VR applications, users are allowed to navigate in virtual scenarios freely. Therefore, in our test, observers can walk freely in the room to watch 3D point clouds, which is different from the way of watching images or videos that subjects only stand or sit in a fixed position. Fig.~\ref{fig:screenshot} shows views of our experimental system from different angles using the HMD.

It is known that distance influences perceptual quality, as observers are more sensitive to errors from a close distance and focus more on the whole from a long distance. If a subject always stands too close to point clouds, then the local views of sparse 3D points are perceived, deranging the test's quality evaluation. So we set a recommended distance of two meters away from models. We suggested the subjects to stand at the distance initially and told them that they could walk freely to navigate and perceive. Eventually, subjects were asked to come back to the initial position, i.e., the fixed position of a scoreboard, to give scores so that the last view would not bias a subject towards a score.
Notably, we made observers walk physically in our experiment instead of teleporting them to a place by controller pointer because observers are more likely to suffer sickness or nausea when the perceptual views switch frequently and are inconsistent with their body movements.

\subsection{Subjective Evaluation Methodology}
Two protocols for subjective PCQA are widely adopted by researchers, namely ACR methodology and DSIS methodology, as shown in Table~\ref{tbl:related work}. 
Subjects tend to rate explicitly according to the visual quality under ACR and relative differences of perception under DSIS. But the double stimulus methodology, DSIS, is more consistent in terms of identifying the level of impairment~\cite{alexiou_performance_2017}. The reason is that point clouds are displayed as a collection of points, and spaces between contents may be habitually recognized as ``holes". Subjects are likely to give lower scores when the reference point cloud is absent. 
Besides, it is mentioned in the recently published standardization ITU-T P.919~\cite{itu919} for immersive $360^{\circ}$ video on HMD that DSIS is statically more reliable than ACR. 

Therefore, our experiment used the DSIS methodology and displayed the reference and the distorted point clouds side by side as shown in Fig.~\ref{fig:screenshot}. In addition, the hidden reference for each sequence was added like some point cloud assessment experiments~\cite{alexiou_towards_2017,torlig_novel_2018,alexiou_comprehensive_2019,alexiou_performance_2017,alexiou_impact_2018,alexious_point_2018,alexiou_point_2018-2,da_silva_cruz_point_2019,9191308}. In other words, a pair of reference point clouds would be simultaneously displayed once for each sequence without the distorted point cloud. Finally, scores from continuous 5-scale rating were then normalized to integer values between 0 and 100~\cite{zerman_subjective_2019,su_perceptual_2019}. It should be noted that side-by-side displaying is commonly used in subjective quality assessment of holographic data like light fields~\cite{8010398,perra2018assessing} and point clouds~\cite{alexiou_towards_2017,torlig_novel_2018,alexiou_comprehensive_2019,alexiou_performance_2017,alexiou_impact_2018,alexious_point_2018,alexiou_point_2018-2,da_silva_cruz_point_2019,zerman_subjective_2019,9191308}. Moreover, the addition of hidden reference records the subjective scores for reference point cloud sequences to reduce the effect of sequence content.

Thirty-eight subjects, aged from 22 to 35, were involved in each session of our subjective test. Seven are experts in video compression or quality assessment. According to the contents, our experiment was separated into two sessions of human figures and inanimate objects. And 22 males and 16 females participated in the first session, while 27 males and 11 females participated in  the second session. Furthermore, 29 subjects engaged in both sessions. 
All subjects were instructed with the procedures, operation, and some attention points of our test during the test guidance. Before the primary test, a test of color blindness was performed, and a training phase was conducted for each subject through the same procedure as the primary test but using an extra sequence \textit{queen}. 
Every session takes about 1 hour with a mandatory break of 5 to 10 minutes to avoid fatigue, presenting distorted versions in random order. 

Compared with SJTU-PCQA~\cite{9238424} and other databases~\cite{9089539}~\cite{9123121}, our database has 20 point cloud sequences, one type of distortion, and 17 distorted levels, which is diverse in sequences and distorted levels. And our database explores the visual quality on HMD under the 6DoF viewing condition.

\section{Data Processing and Analysis}\label{section4}
In this section, Differential Mean Opinion Scores (DMOS) are calculated as final scores after outlier detection of subjects and samples. The correlation between human figure session and inanimate object session is analyzed. Then, the impacts of different sequences and texture/geometry QPs are discussed.

\subsection{Outlier Detection \& DMOS}



For comparisons with point clouds captured from different methods, the raw scores were first converted to the difference score between the reference and the distorted point clouds, 
\begin{equation}
d_{i,j} = s_{ref_{i,j}} - s_{i,j},
\end{equation}
where $s_{ref_{i,j}}$ and $s_{i,j}$ respectively stand for the subjective rating of the original and distorted point clouds from subject $i$ over samples of the point cloud $j$, where $i \in [ 1,\cdots,N]$ and $j \in [ 1,\cdots,M]$ with $N$ and $M$ being the numbers of subjects and samples, respectively. $d_{i,j}$ is the difference score of subject $i$ rating on sample $j$. The score from a subject over all samples is represented as $\mathbf{s_i}=(s_{i,1},\cdots,s_{i,M})$. 
An outlier detection procedure from ITU-R BT 500.13 recommendation~\cite{itu500} was then used to remove scores generated by unreliable subjects. 
In our study, none of subjects were rejected in the procedure.

In order to unify the different rating scales across subjects, difference scores for each observer were transformed to z-scores
 like
~\cite{1284395}
 by the mean and the standard deviation equal for all observers,
\begin{equation} 
    z_{i,j} = \frac{d_{i,j} - \bar{d}_{i}}{\sigma_{i}},
\end{equation} 
where $\bar{d}_i$ and $\sigma_i$ show the mean value and standard deviation of a subject. 
Then scores assigned by a subject would be normalized to zero mean and unit variance. It is known that $99\%$ of values fall within three standard deviations from the zero mean within a normally distributed sample. Therefore, z-scores are rescaled by linear mapping~\cite{5404314},
\begin{equation} 
    \hat{z}_{i,j}=\frac{z_{i,j}+3}{6}.
\end{equation} 
Finally, DMOS of each test point cloud is computed as the mean of the rescaled z-scores as
\begin{equation} 
    DMOS_j = \frac{1}{N}\sum_{i=1}^{N} \hat{z}_{i,j}.
\end{equation} 
In particular, screening of observers and the score processing procedure are accomplished for each session.

\subsection{Correlation between Two Sessions}

\begin{figure}
  \centerline{\includegraphics[width=0.8\linewidth]{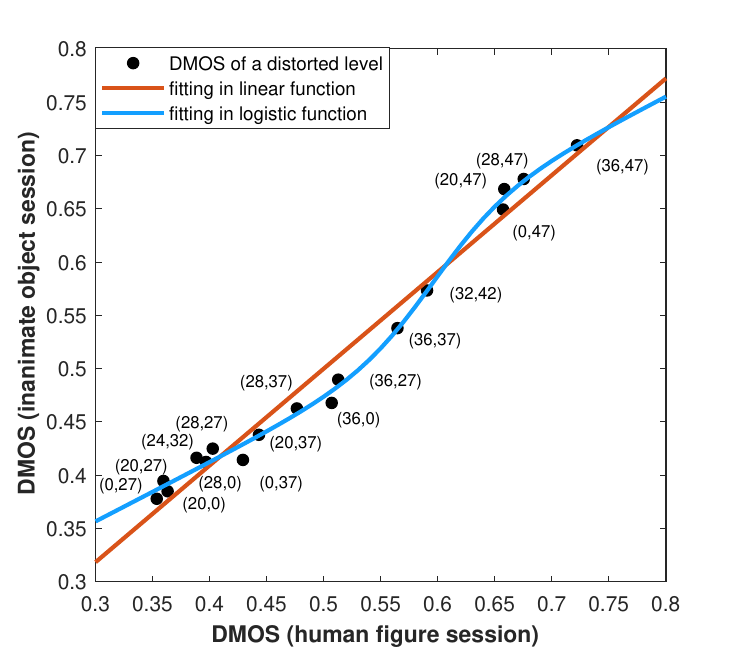}}
  \caption{Correlation between the human figure session and inanimate object session.}
  \label{fig:two_sessions}
\end{figure}

\begin{figure}
    \centering
    \subfigure[]{
        \centering
        \includegraphics[width=1\columnwidth]{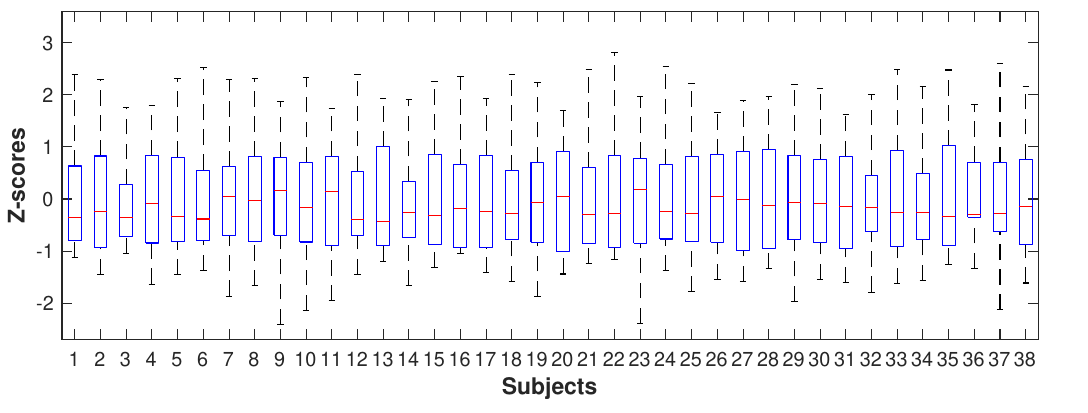}
        \label{fig:session1}
    }
    \subfigure[]{
        \centering
        \includegraphics[width=1\columnwidth]{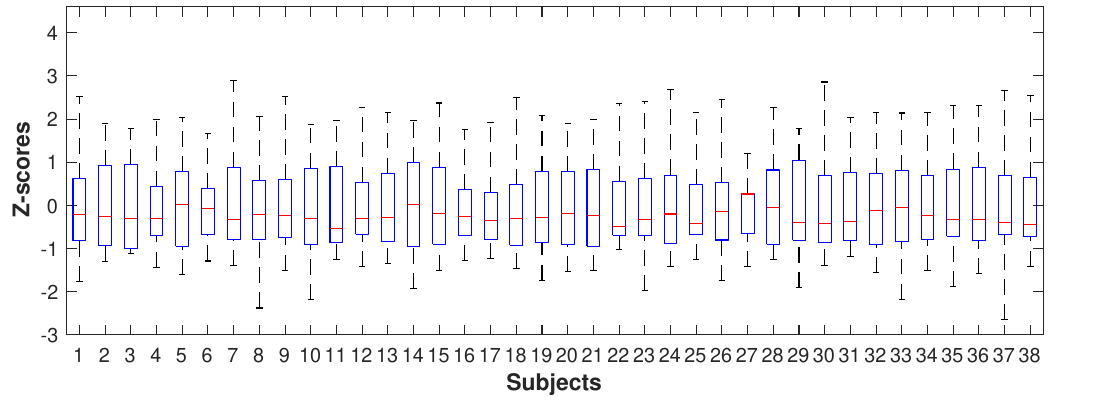}
        \label{fig:session2}
    }
    \caption{The boxplot of z-scores in two sessions. (a) the human ﬁgure session, (b) the inanimate object session.
    }
    \label{fig:sessions}
\end{figure}

To avoid fatigue, our tests were separated into two parts by considering different categories of contents, i.e., human figures and inanimate objects. We examine the correlation between the DMOSs of these two sessions and use a linear function and a logistics function to fit the data. As shown in Fig.~\ref{fig:two_sessions}, a black point denotes a rate level, shown as (geometry QP, texture QP). The R-square value of linear fitting and logistic fitting are 0.973 and 0.9628. It indicates the linear correlation of the two sessions in our test, which means that there is no significant difference caused by different subjects and processes. It can be seen that DMOSs increase when geometry and texture QPs are enlarged, and the degradation of point clouds compressed by QP of medium values are more visually distinguishable. 
In addition, the boxplot of z-scores of subjects in both sessions is shown in Fig.~\ref{fig:sessions}. 
Z-scores data have stable distributions in both session 1 and session 2 with the minimum and maximum values of -2.4095 and 2.8213 in session 1 and of -2.6614 and 2.8863 in session 2.

\subsection{Analysis}

\begin{figure*}
    \centering
    \subfigure[]{
        \centering
        \includegraphics[width=0.4\textwidth]{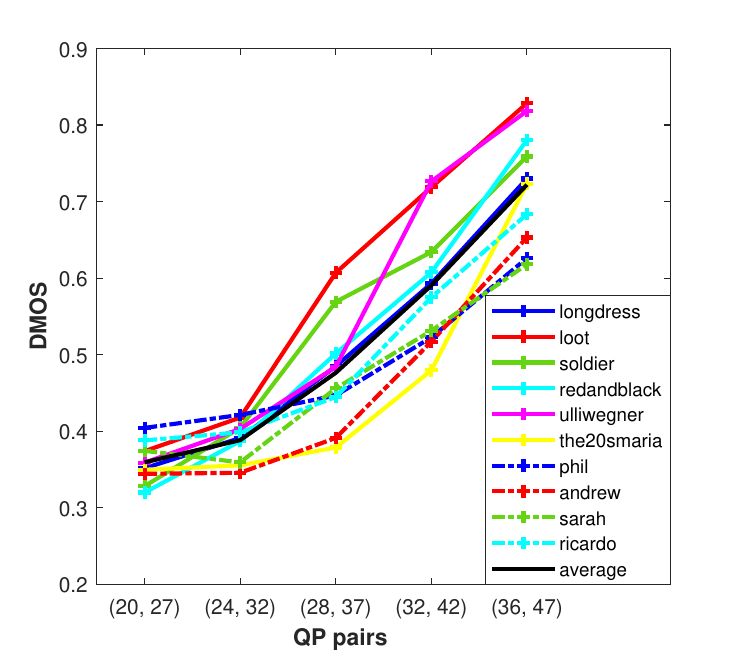}
        \label{fig:CTC1}
    }
    \subfigure[]{
        \centering 
        \includegraphics[width=0.4\textwidth]{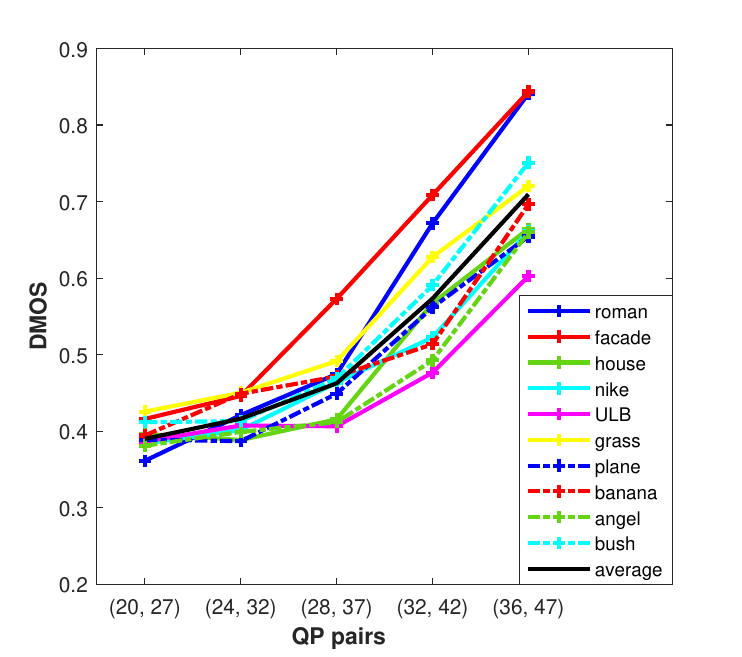}
        \label{fig:CTC2}
    }
    \caption{Comparing the visual quality of different sequences under QP pairs defined in CTC from MPEG PCC. (a) the human figure session, (b) the inanimate object session. 
    } 
    \label{fig:CTC}
    \end{figure*}

In Fig.~\ref{fig:CTC}, we compare the visual quality of different sequences under QP pairs defined in CTC from MPEG PCC. 
As can be seen in all sequences, there is slight visual quality degradation from $(20,27)$ to $(24,32)$. Observers are sensitive to the compression distortion when QP pairs arise from $(28,37)$ to $(36,47)$. 
Besides, the lower values and gentle changes, denoted as dotted lines in Fig.~\ref{fig:CTC1}, indicate that it is harder for subjects to perceive the distortions in the upper body figure set due to the quality of the source point clouds.

\begin{figure}
    \centering
    \subfigure[]{
        \includegraphics[width=0.45\columnwidth]{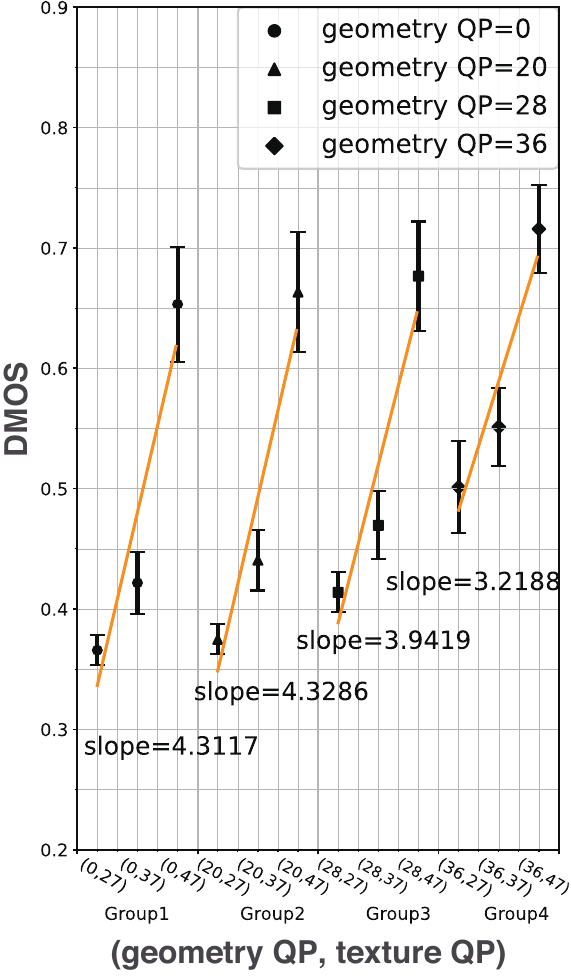}
        \label{fig:geometry}
    }
    \subfigure[]{
        \includegraphics[width=0.45\columnwidth]{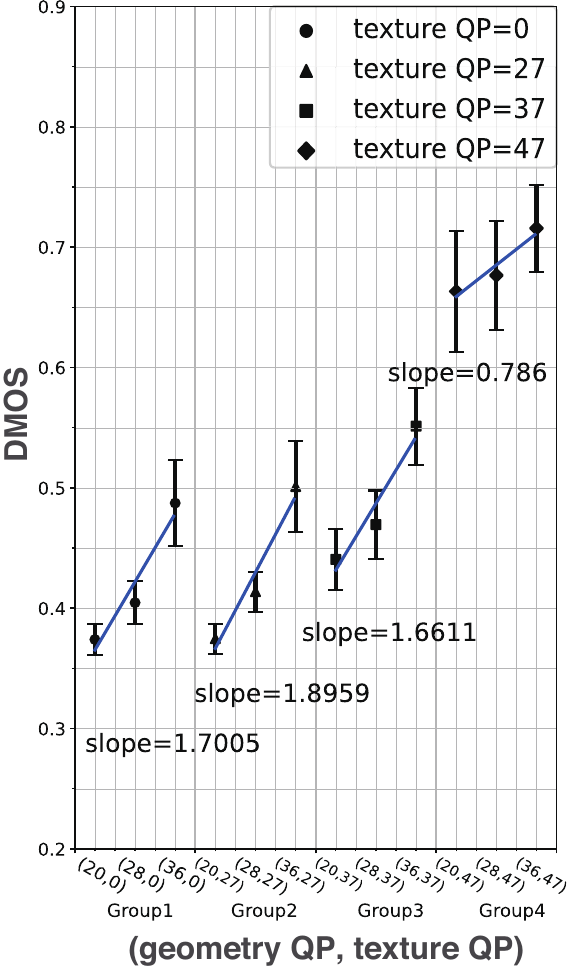}
        \label{fig:texture}
    }
    \caption{Varying geometry/texture QP against subjective scores. (a) grouping by geometry QP, (b) grouping by texture QP.
    }
    \label{fig:geo_text_2d}
\end{figure}

Fig.~\ref{fig:geo_text_2d} shows varying geometry and texture QPs against subjective scores. Varying texture QPs can be seen in each group of Fig.~\ref{fig:geometry}, and varying geometry QPs can be seen in each group of Fig.~\ref{fig:texture}. 
Three findings in Fig.~\ref{fig:geo_text_2d} are listed as follows. 
\textit{1)} Group 3 and Group 4 in Fig.~\ref{fig:texture} represent the texture QP values of 37 and 47, respectively. It can be seen that the values in Group 4 are much higher than values in Group 3, meaning that perceptual quality falls off with texture QP rising from 37 to 47. 
\textit{2)} The slope of Group 4 in Fig.~\ref{fig:texture} is smaller than other groups in Fig.~\ref{fig:texture}. In other words, when texture QP equals 47 and geometry QP is between 20 and 36, the variation of subjective scores within the group of texture QP as 47 is relatively smaller than groups of other texture QP values like 0, 27, and 37. 
\textit{3)} In addition, it can be seen that the slopes in Fig.~\ref{fig:geometry} are higher than slopes in Fig.~\ref{fig:texture}.

\section{Proposed Projection-based objective Models}\label{section5}

Based on our subjective PCQA database, we evaluate the performance of popular point-based objective methods used by MPEG, as shown in Table~\ref{tbl:performance_3D}. The bold denotes the best performance of a correlation coefficient, i.e., a column in Table~\ref{tbl:performance_3D}. 
In particular, predictive objective scores are obtained through nonlinear regression, according to the work~\cite{1576816}. 
Then, based on the Recommendation ITU-T P.1401~\cite{itu1401}, performance evaluation metrics are adopted to show the performance, including Pearson's Linear Correlation Coefficient (PLCC), Spearman Rank Order Correlation Coefficient (SROCC), Kendall Rank Order Correlation Coefficient (KROCC), and Root Mean Square Error (RMSE). 



It can be found that objective scores show a low correlation with subjective scores. D1 MSE reaches 0.3136 in PLCC and 0.3963 in SROCC, while D2 MSE shows correlation as 0.3498 in PLCC and 0.4125 in SROCC. The reason may be that D1 and D2 only consider the geometry and ignore color information. Moreover, D1/D2 and YUV color are error-based models without taking characteristics of human eyes into account, although YUV color, which denotes comparison of texture attributes after point matching, can slightly better represent visual quality compare to geometry attributes (D1/D2). 
Therefore, we desire to predict the visual quality of point clouds by mimicking how people perceive the world. 



\begin{table*}
\caption{Performance of point-based objective point cloud quality metrics.}
\begin{center}
\resizebox{\textwidth}{!}{
\begin{tabular}{c|l|c c c c|c c c c | c c c c }
\hline
\multirow{2}{*}{Category}&\multirow{2}{*}{Metric} & \multicolumn{4}{c|}{All} & \multicolumn{4}{c|}{Human figure session} & \multicolumn{4}{c}{Inanimate object session} \\ 
&& PLCC & SROCC & KROCC & RMSE & PLCC & SROCC & KROCC & RMSE & PLCC & SROCC & KROCC & RMSE \\ \hline

\multirow{4}{*}{D1~\cite{N17995}}&p2point MSE &
 0.3136 & 0.3963 & 0.2761 & 0.1224 & 0.3850 & 0.4553 & 0.3299 & 0.1221 & 0.2549 & 0.3517 & 0.2459 & 0.1213 
 \\ 
&p2point Hausdorff & 
 0.2980 & 0.3791 & 0.2620 & 0.1231 & 0.3823 & 0.4514 & 0.3261 & 0.1223 & 0.2311 & 0.3215 & 0.2228 & 0.1220 
 \\ 
&PSNR-p2point MSE & 
0.2849 & 0.3279 & 0.2252 & 0.1236 & 0.3103 & 0.3957 & 0.2770 & 0.1258 & 0.3205 & 0.2794 & 0.1898 & 0.1188
 \\ 
&PSNR-p2point Hausdorff & 
0.2825 & 0.3273 & 0.2268 & 0.1237 & 0.2996 & 0.3894 & 0.2745 & 0.1262 & 0.3178 & 0.2827 & 0.1940 & 0.1189 
 \\ 
\hline
\multirow{4}{*}{D2~\cite{tian_geometric_2017}}&p2plane MSE & 
0.3498 & 0.4125 & 0.2947 & 0.1208 & 0.4151 & 0.4759 & 0.3548 & 0.1204 & 0.2971 & 0.3542 & 0.2504 & 0.1198
\\ 
&p2plane Hausdorff & 
0.3218 & 0.3862 & 0.2679 & 0.1221 & 0.3993 & 0.4622 & 0.3378 & 0.1213 & 0.2323 & 0.3138 & 0.2135 & 0.1220 
\\ 
&PSNR-p2plane MSE  & 
0.3111 & 0.3463 & 0.2381 & 0.1225 & 0.3631 & 0.4371 & 0.3069 & 0.1233 & 0.2906 & 0.2781 & 0.1840 & 0.1200
 \\ 
&PSNR-p2plane Hausdorff &
0.3013 & 0.3419 & 0.2363 & 0.1229 & 0.3471 & 0.4039 & 0.2826 & 0.1241 & 0.3177 & 0.2933 & 0.2006 & 0.1189 
\\ 
\hline
\multirow{4}{*}{YUV~\cite{N17995}}&PSNR-Y & 
0.3443 & 0.3481 & 0.2318 & 0.1211 & \textbf{0.5920} & \textbf{0.5295} & \textbf{0.3609} & \textbf{0.1067} & 0.3991 & 0.4338 & 0.2945 & 0.1150 
\\ 
&PSNR-U & 
0.3883 & 0.4097 & 0.2790 & 0.1188 & 0.4902 & 0.4527 & 0.3079 & 0.1153 & 0.4549 & 0.5195 & 0.3653 & 0.1117 
 \\ 
&PSNR-V & 
\textbf{0.4373} & 0.4378 & 0.3035 & \textbf{0.1160} & 0.4502 & 0.4244 & 0.3007 & 0.1182 & 0.4677 & 0.4840 & 0.3342 & 0.1109 
\\ 
&PSNR-YUV & 
0.4336 & \textbf{0.4544} & \textbf{0.3098} & 0.1162 & 0.5230 & 0.5058 & 0.3507 & 0.1128 & \textbf{0.4794} & \textbf{0.5417} & \textbf{0.3799} & \textbf{0.1101} 
 \\
\hline
\end{tabular}
}
\end{center}
\label{tbl:performance_3D}
\end{table*}
 
\subsection{Proposed Weighted View Projection Based PCQA Method}
\begin{figure}
\centerline{\includegraphics[width=1\columnwidth]{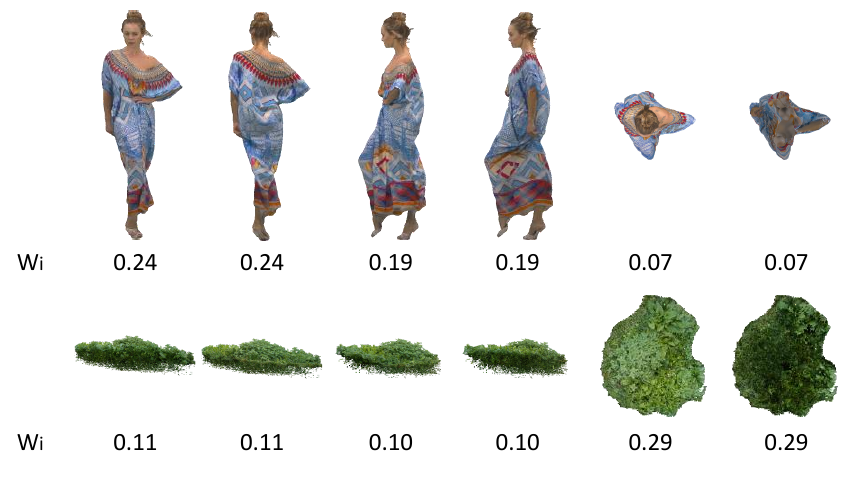}}
\caption{Six view images through orthographic projection obtained from the sequence \textit{Longdress} (the first row) and the sequence \textit{Grass} (the second row). The view images, in order, are the front, back, left, right, top, and bottom views.}
\label{fig:view_projection}
\end{figure}

The widely used projection-based method~\cite{torlig_novel_2018} projects a point cloud onto six planes of the bounding box, viewing each plane as an equal role regardless of the significance of different views. 
Through the subjective experiment, we found that the size of a plane of bounding box might relate to visual quality as the larger region own a bigger chance to capture subjects' visual attention by showing more details of the content. For instance, observers show a tendency of ignoring the top and bottom view of a human digital sequence as well as spending more time in various views in the object models, such as the front view of sequence \textit{ULB\_Unicom}, the top view of sequence \textit{grass} and side views of the sequence \textit{Nike}. 
Examples of the projected images obtained from the sequence \textit{longdress} and \textit{grass} are exhibited in Fig.~\ref{fig:view_projection}.

\begin{figure*}[htb]
    \centerline{\includegraphics[width=1\linewidth]{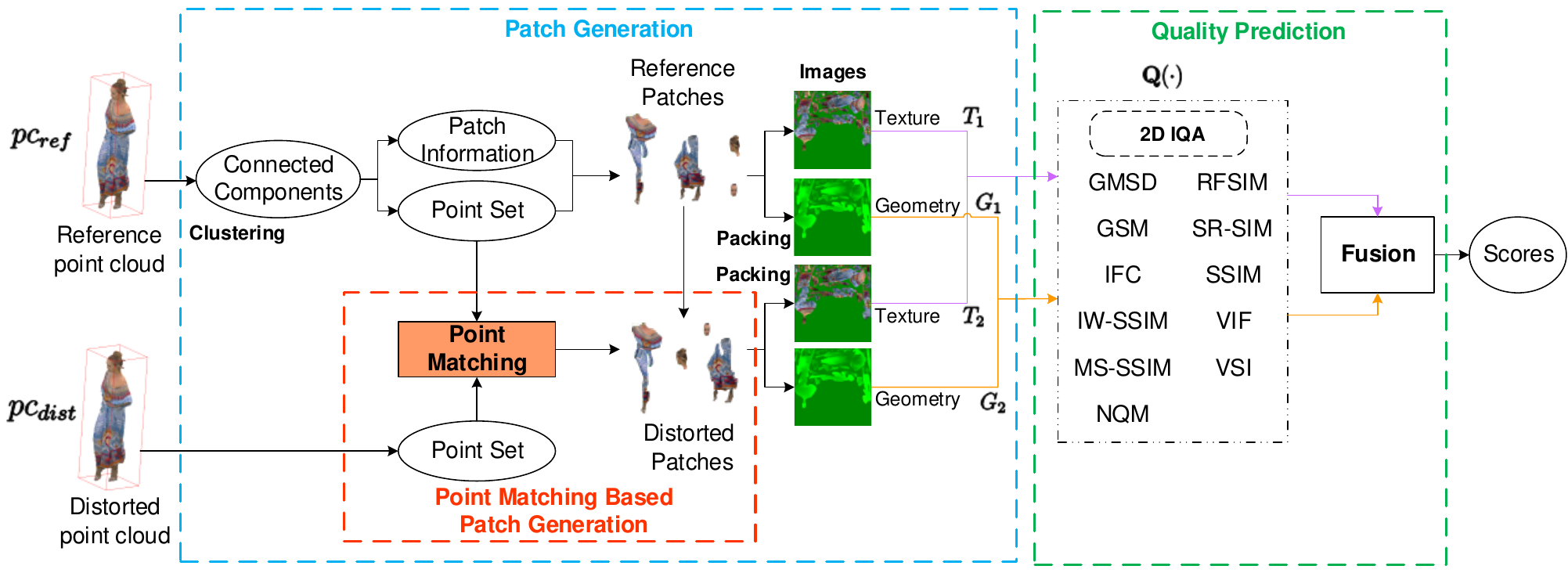}}
    \caption{The framework of the proposed patch projection based PCQA method.}
    \label{fig:patch_projection_process}
    \end{figure*}
 
The process of view projection based model can be formulated as 
\begin{equation}
    \begin{aligned}
    S_{final} = \sum_{i=1}^{6}w_{i}\cdot \mathbf{Q}(\mathbf{P_i}(pc_{ref}), \mathbf{P_i}(pc_{dist})),
    \end{aligned}
\end{equation}
where $S_{final}$ denotes the objective scores of this projection-based metric. $\mathbf{P_i}(\cdot)$ indicates projecting a reference point cloud $pc_{ref}$ or a distorted point cloud $pc_{dist}$ onto a bounding box plane, and $i\in [1,...,6]$ corresponds to the front, back, left, right, top, and bottom views, respectively. $\mathbf{Q}(\cdot)$ denotes the operation of computing visual quality scores using one of the existing IQA methods. 

Here, we propose a weighted view projection based PCQA method by setting the weight of one view $w_i$ as the ratio of the size of a plane to the sum  of the area of six planes on the bounding box. It is consistent with the viewing habits of subjects for the reason that point clouds are characterized by keeping the content in the center, with the subject moving around, and the larger view has larger chance to be seen. So $w_i$ is computed as $\frac{c_i}{\sum_{i=1}^{6}{c_{i}}}$, and $c_i$ is the area of a plane of the bounding box.

This model is easy to operate, time-saving, and computationally cheap. However, view projection may result in occlusion, and geometry may be less sensitive or incompletely expressed, as observers can see a point cloud from every angle, and six planes are inadequate to denote all views sensed in human eyes.

\subsection{Proposed Patch Projection Based PCQA Method}

To discover more details of different views for a point cloud, it is better to segment a point cloud into smaller parts and then project them onto planes. A smaller connected part, namely a patch, is composed of 3D points with similar normal vectors. To reduce the self-occlusion brought by view projection, we propose an objective PCQA model based on patch projection, as shown in Fig.~\ref{fig:patch_projection_process}. The whole process has two parts, i.e., patch generation and quality prediction. In patch generation, the reference and the distorted point clouds are converted into two geometry images and two texture images, each of which includes non-overlapping patches. In quality prediction, the visual quality of geometry and texture images are computed to denote the visual quality of the distorted point cloud.

\begin{figure}
    \centering \includegraphics[width=1\linewidth]{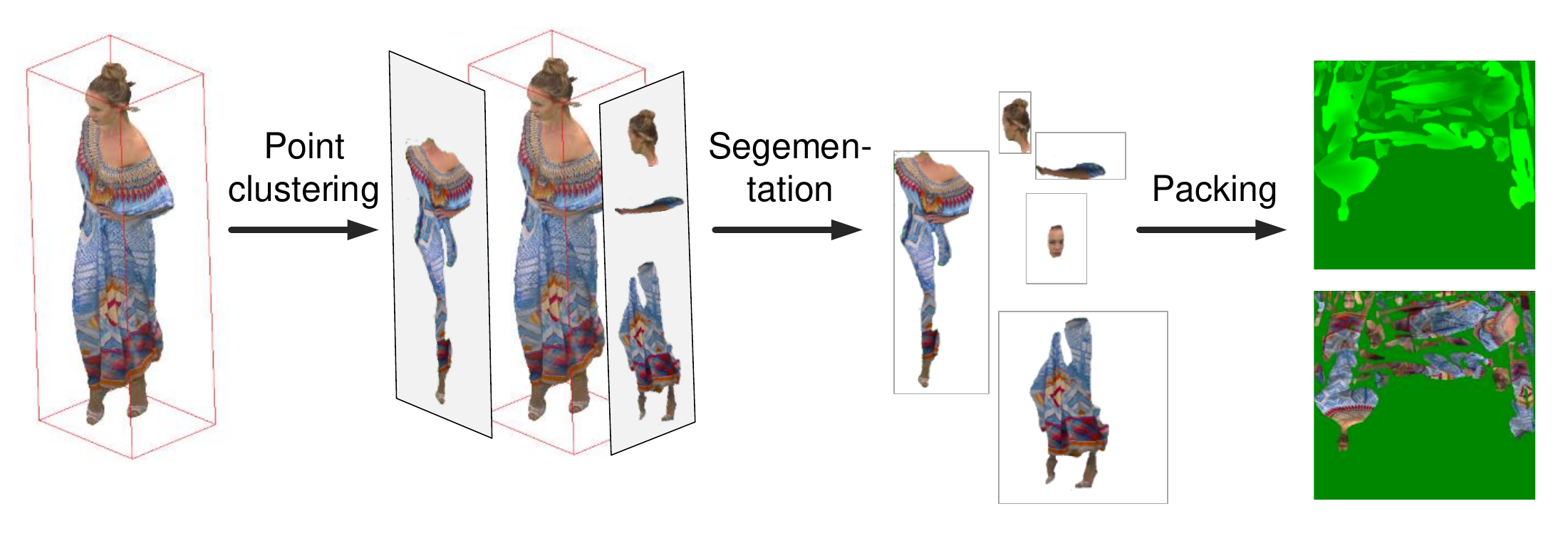}
    \caption{The process of 3D to 2D patch projection for a point cloud.}  
    \label{fig:patch_projection_one_PC}
\end{figure}

Fig.~\ref{fig:patch_projection_one_PC} depicts the process of 3D to 2D patch projection for a reference point cloud. For the reference point cloud in patch generation, patches can be obtained by clustering 3D points according to the direction of normal vectors and segmenting the connected components. Then all patches are inserted into a blank image gird in the packing process to create one geometry image and one texture image.

\begin{figure}
    \centering
          \subfigure[]{
            \begin{minipage}[b]{0.295\linewidth}
            \includegraphics[width=1\linewidth]{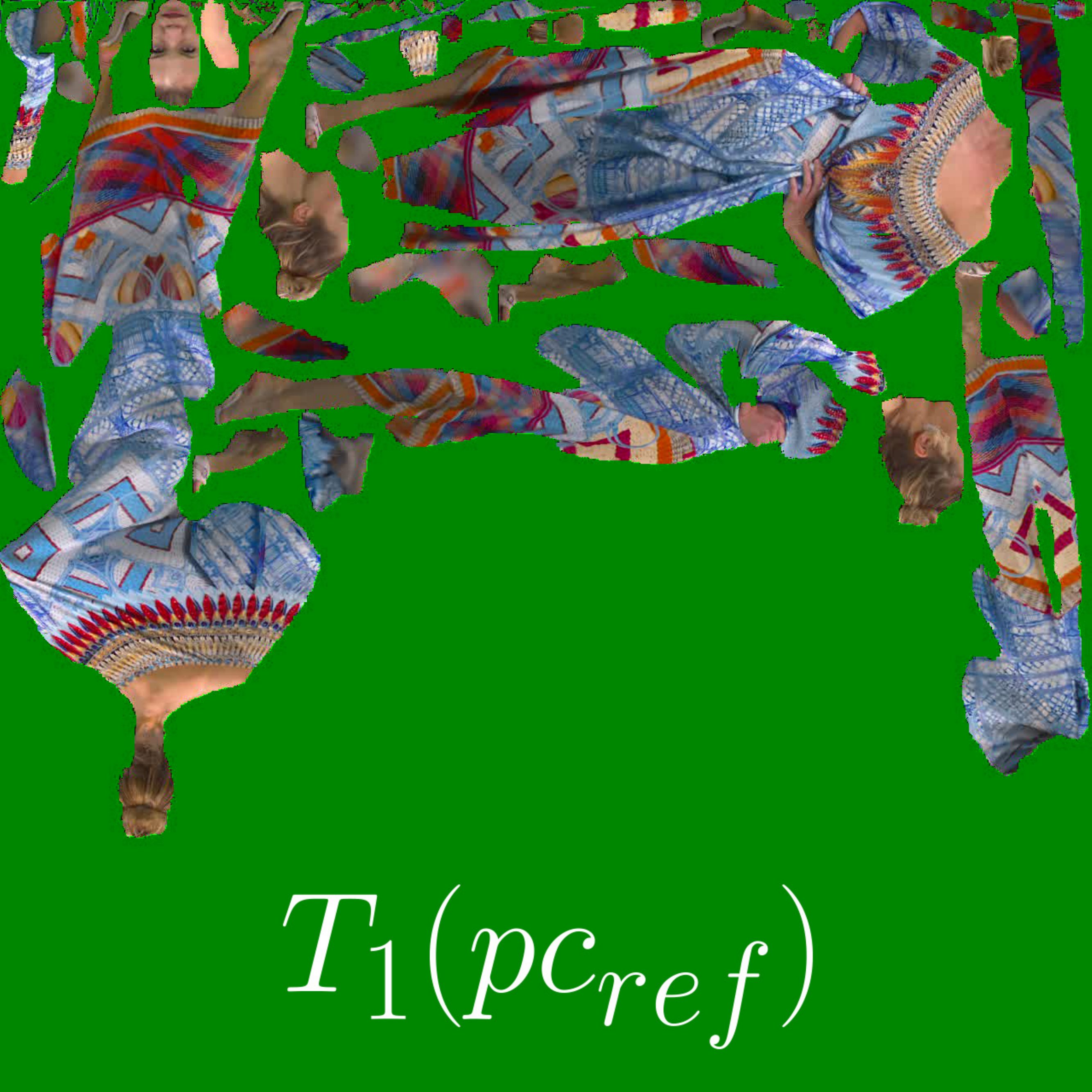}\vspace{4pt}
            \includegraphics[width=1\linewidth]{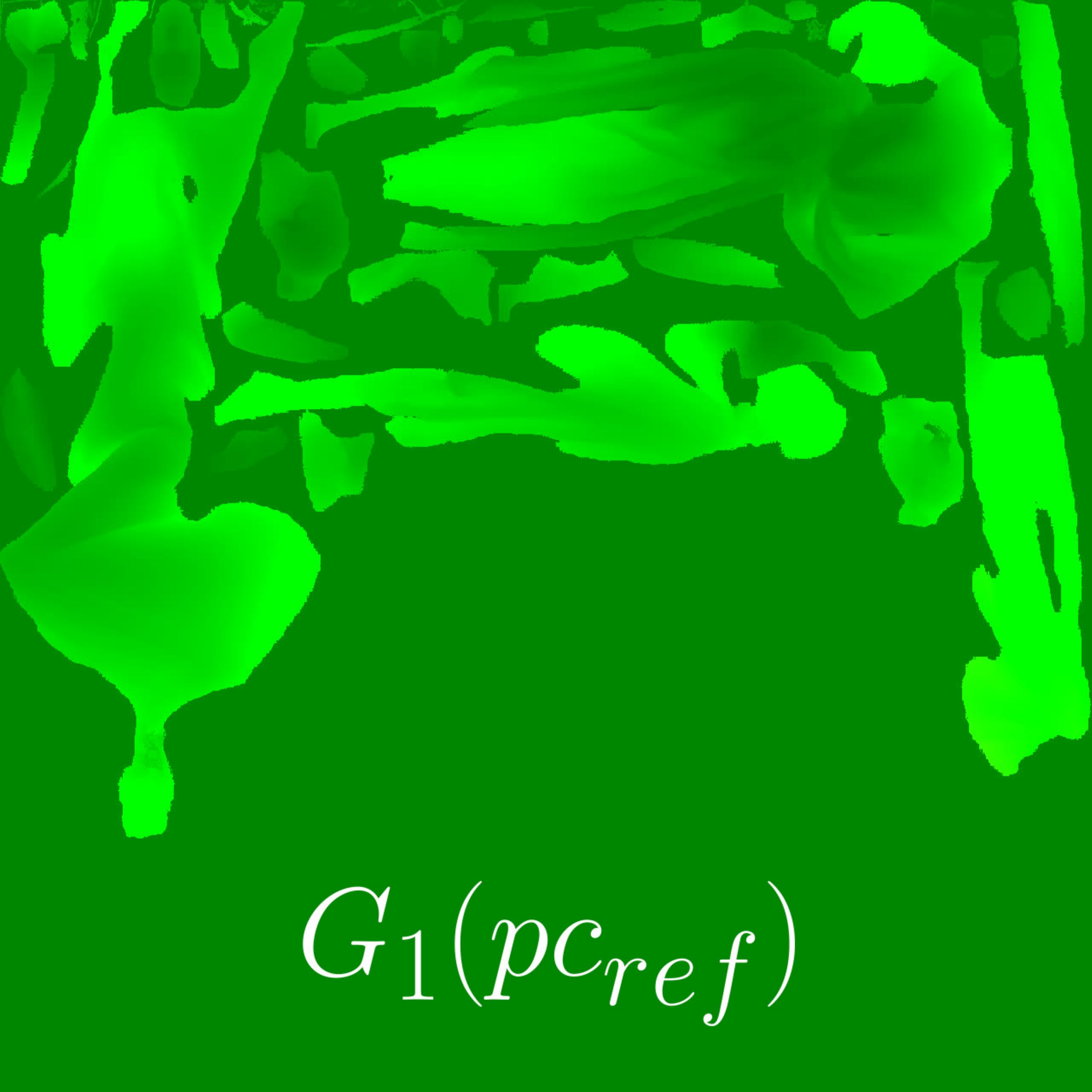}
            \end{minipage}
          }
          \subfigure[]{
            \begin{minipage}[b]{0.295\linewidth}
            \includegraphics[width=1\linewidth]{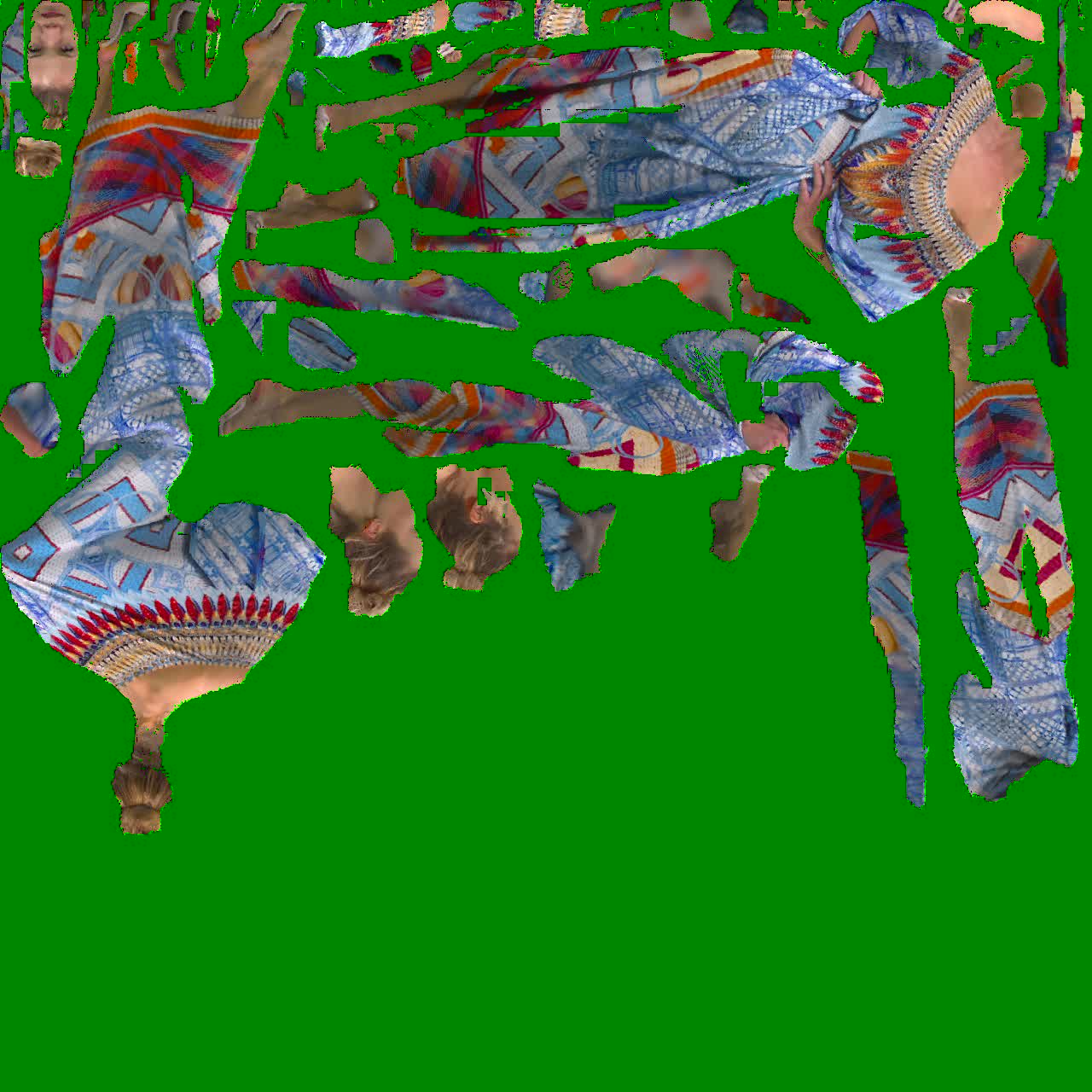}\vspace{4pt}
            \includegraphics[width=1\linewidth]{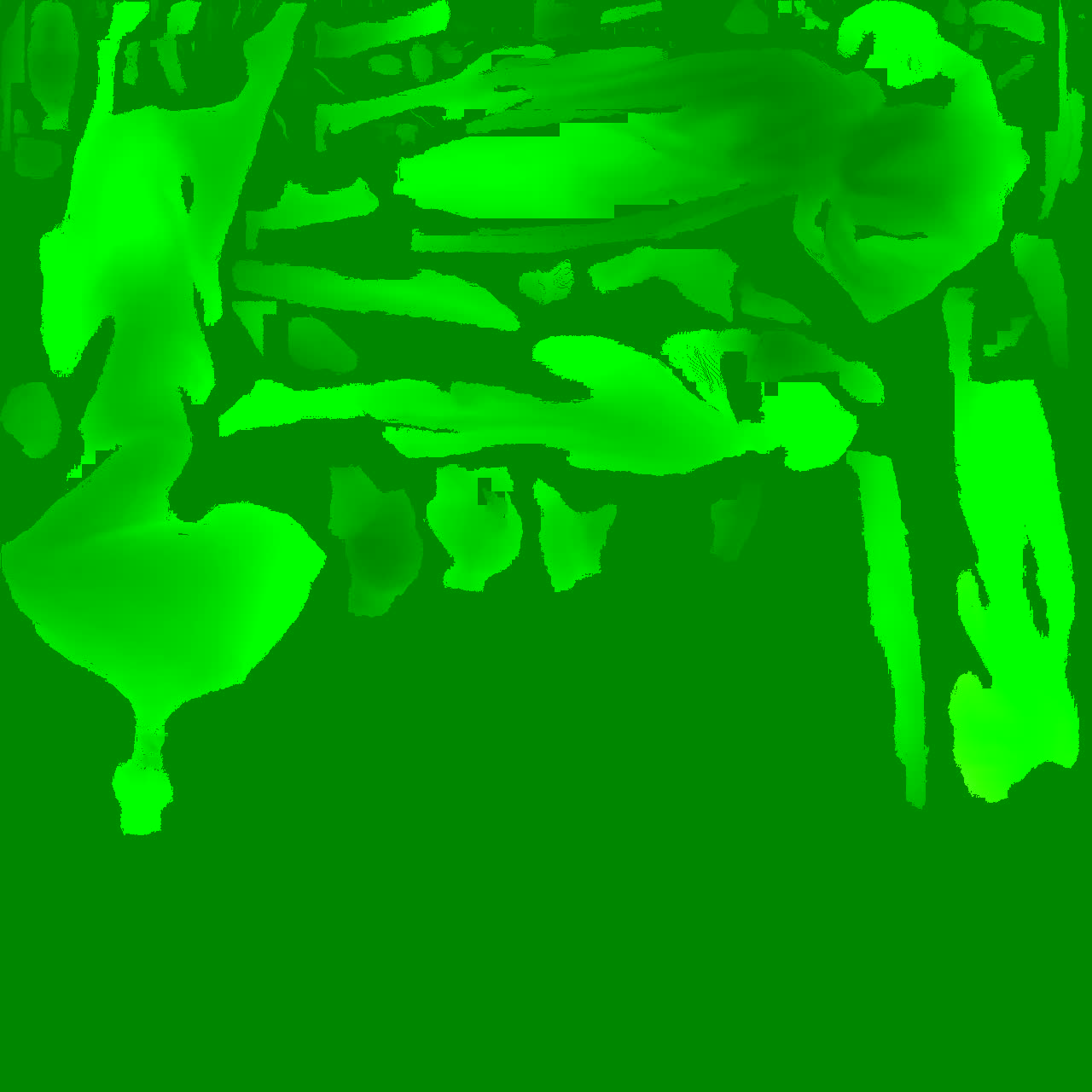}
            \end{minipage}
          }
          \subfigure[]{
            \begin{minipage}[b]{0.295\linewidth}
            \includegraphics[width=1\linewidth]{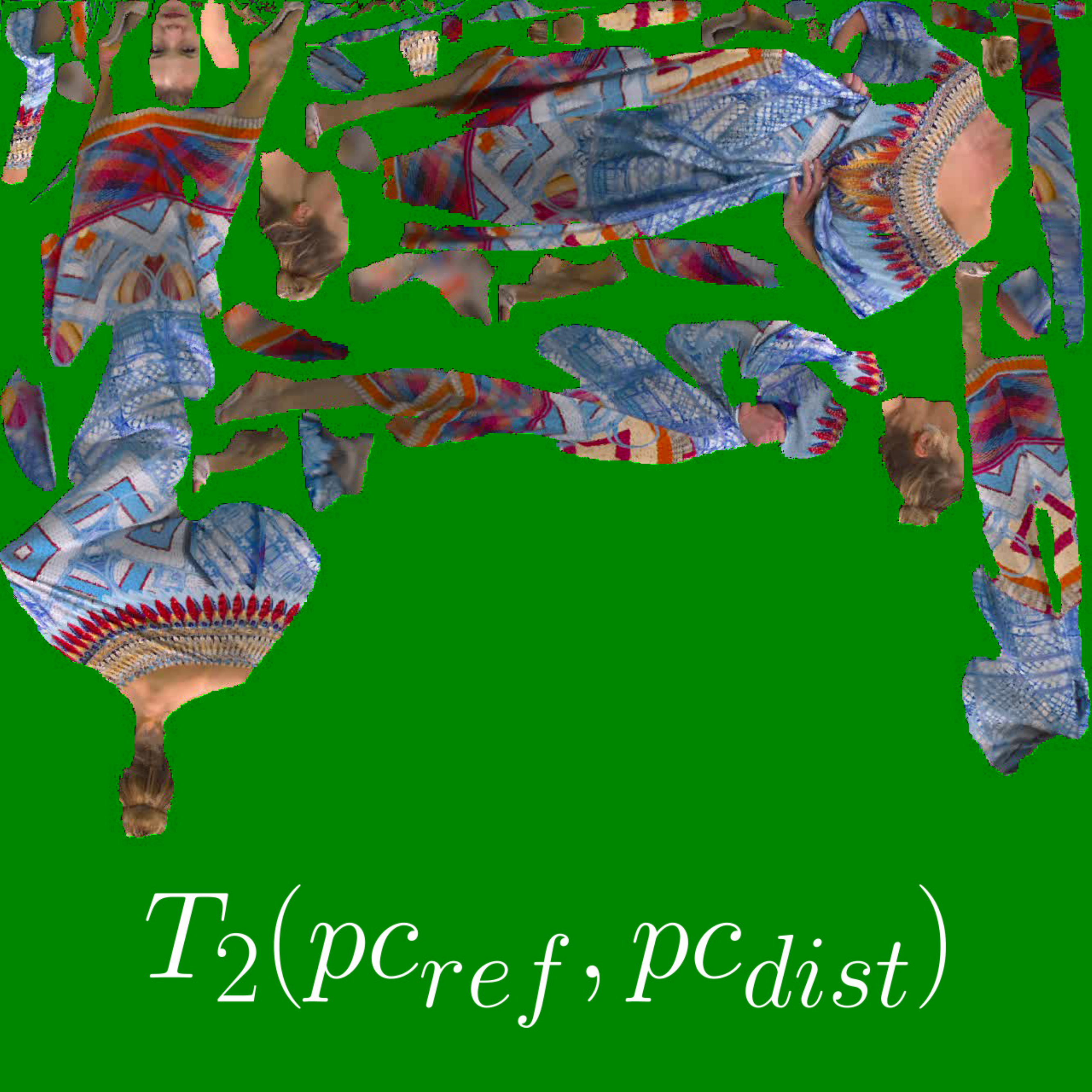}\vspace{4pt}
            \includegraphics[width=1\linewidth]{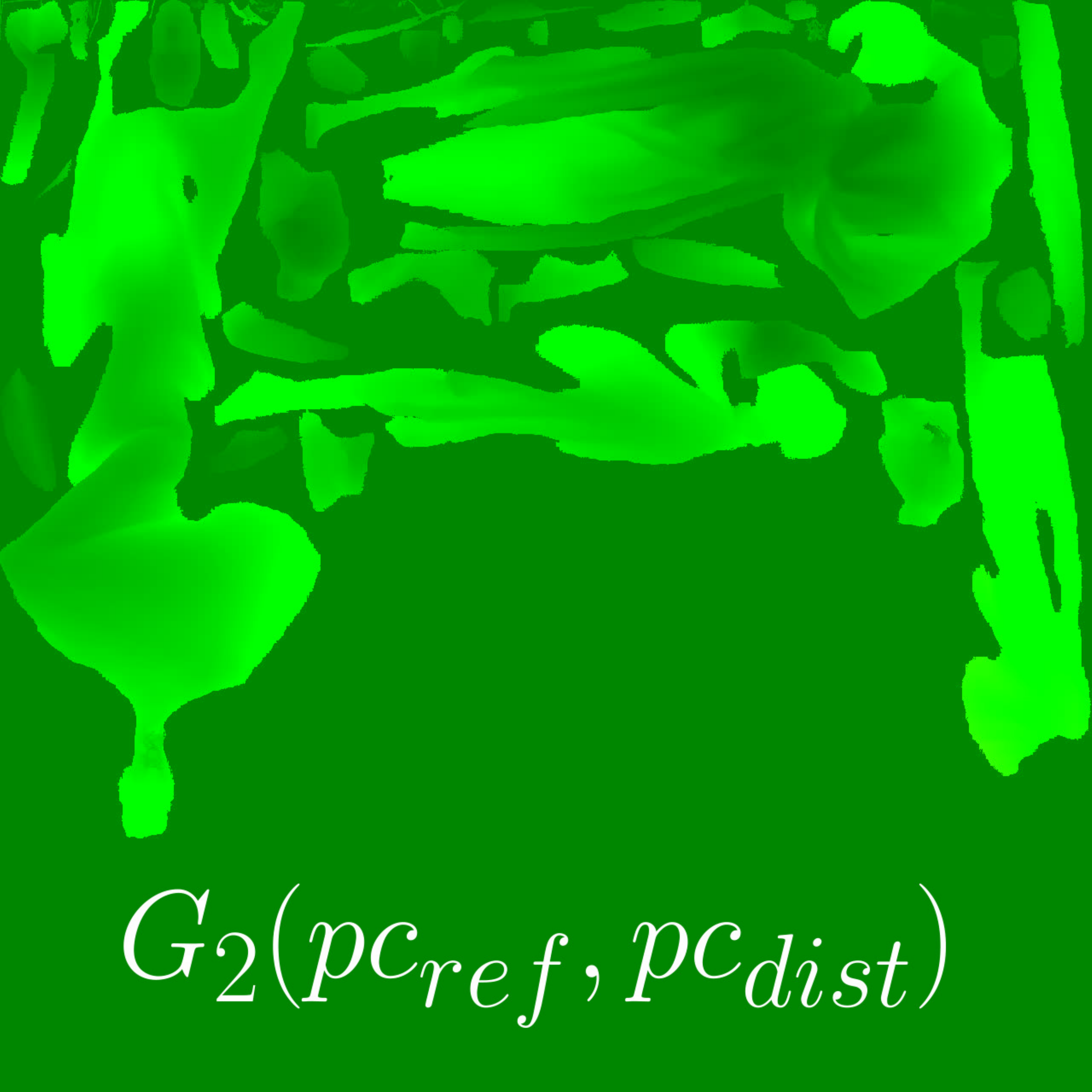}
            \end{minipage}
          }
    \caption{Texture/geometry images obtained from the reference and the distorted point clouds. (a) from the reference point cloud, (b) from the distorted point cloud without point matching, (c) from the distorted point cloud with point matching.}  
    \label{fig:patch_projection}
\end{figure}

However, since the geometry distortion may change the positions and normal vectors, the patches of the distorted point cloud generated from the procedure in Fig.~\ref{fig:patch_projection_one_PC} may differ from those of the reference point cloud. The first and the second column in Fig.~\ref{fig:patch_projection} shows one texture image and one geometry image obtained from the reference and the distorted point clouds, respectively. We can observe that there are significant mismatches between the images generated from the reference and the distorted point clouds. These mismatches cause the conventional full-reference 2D IQA methods are no longer applicable to geometry and texture images from the distorted point cloud.

To handle this mismatch problem, we design a point matching based patch generation for the distorted point cloud. Algorithm 1 describes the point matching based patch generation algorithm. 
At first, the corresponding points in the reference and the distorted point clouds are found by the nearest neighbor algorithm. Then, new patches are assigned by reference patches, and then each point in these new patches is replaced by its corresponding point in the distorted point cloud. In a word, new patches maintain the contours and placement information of reference patches, but each point in new patches is obtained from the distorted point cloud. Finally, new patches are viewed as distorted patches. 
Based on the reference patch information and correspondences between points of reference and distorted point clouds, distorted patches are able to match the reference patches correspondingly. 
The third column in Fig.~\ref{fig:patch_projection} shows the geometry and the texture images generated from the distorted point cloud using Algorithm 1. We can find that the patches in the third column match those in the first column accurately. Consequently, we can use the conventional 2D IQAs to do the quality prediction.

\begin{algorithm}
    \caption{Point matching based patch generation}\label{algorithm}
    \KwData{reference point set $A$, distorted point set $B$, reference patches $T_1$}
    \KwResult{distorted patches $T_2$}
    
    \ForEach{point $p$ $\in$ $A$}{
    $MatchedPoint[p]$ $\leftarrow$ NearestNeighbor($p$, $B$)\;
    }
    
    $new\ patches\ T \leftarrow T_1$\;
    \ForEach{patch $t$ $\in$ $T_2$}{
        \ForEach{point $q$ $\in$ $t$}{
            $q$ $\leftarrow$ $MatchedPoint[q]$;
        }
    }
    $T_2 \leftarrow T$\;
    \label{code}
    \end{algorithm}

In quality prediction, geometry or texture images are evaluated by one of effective IQA metrics respectively, then the geometry and texture quality scores are fused by addition to obtain the final score. 
The process can be formulated as, 
\begin{equation}
\begin{aligned}
S_{final} = a\cdot \mathbf{Q}(\mathbf{T_1}(pc_{ref}), \mathbf{T_2}(pc_{ref}, pc_{dist}))\\+b\cdot \mathbf{Q}(\mathbf{G_1}(pc_{ref}), \mathbf{G_2}(pc_{ref}, pc_{dist})),
\end{aligned}
\end{equation}
where $\mathbf{T_1}(\cdot)$ and $\mathbf{G_1}(\cdot)$ indicate projecting a reference point cloud $pc_{ref}$ and generating a texture image and a geometry image, respectively. Similarly, $\mathbf{T_2}(\cdot)$ and $\mathbf{G_2}(\cdot)$ denote projecting a distorted point cloud $pc_{dist}$ and creating the texture and the geometry images with the same patch placement of $\mathbf{T_1}(\cdot)$ and $\mathbf{G_1}(\cdot)$. 
$\mathbf{Q}(\cdot)$ denotes using one of IQA methods, computing visual quality scores of geometry or texture images. Then, the fusion of geometry and texture images obtains the final score of a point cloud, i.e., $S_{final}$. 
In joint bit allocation between geometry and color for V-PCC~\cite{9194311}, the distortion model for point clouds is modeled as a linear combination of the geometry distortion and texture distortion, and the weighting factor of texture distortion is set as 0.75 or 0.5. Thus, we chose one of median values for the texture parameter between 0.5 and 0.75 as 0.6. 
Here, $a$ and $b$ are set as 0.6 and 0.4, respectively.

\begin{figure}
\centerline{\includegraphics[width=0.65\linewidth]{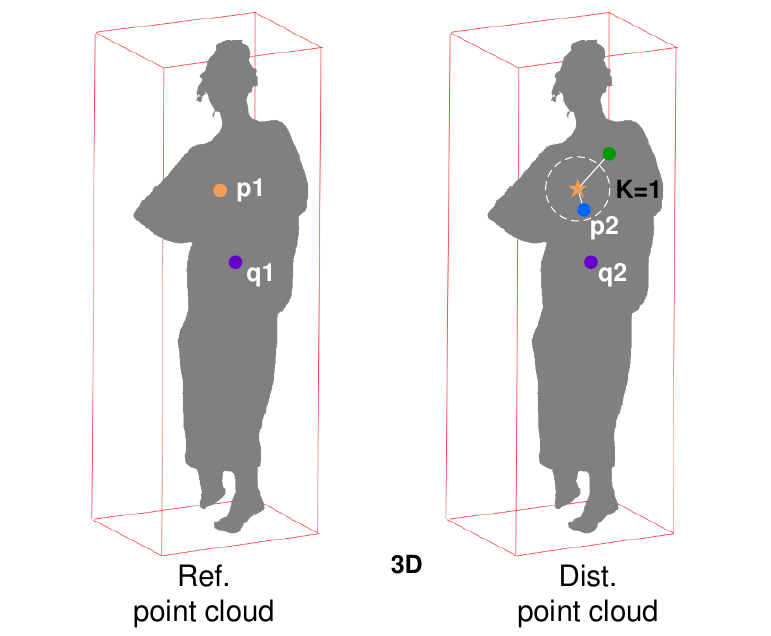}}
\caption{Point matching between reference and distorted point clouds.}
\label{fig:patch_projection_point_correspondence}
\end{figure}

Following V-PCC mechanism, we implemented our patch projection based PCQA method based on TMC2, and changes have been made to generate distorted patches according to the following Algorithm 1. Point matching is achieved by the KNN algorithm, where we set $K=1$ to find the nearest neighbor. 
In the process of point matching, for each point $p_1$ in the reference point cloud, we find the nearest neighbor $p_2$ in the distorted point cloud 
as illustrated in Fig.~\ref{fig:patch_projection_point_correspondence}. 

\section{Experimental Results and Analysis}\label{section6}
To evaluate the performance of the proposed two projection-based PCQA methods, two datasets SIAT-PCQD and vsenseVVDB~\cite{zerman_subjective_2019} were used. In addition to the conventional D1 and D2 metrics in Table~\ref{tbl:performance_3D}, another two kinds of benchmark schemes were compared. One is point-based PCQA metrics, including PC-MSDM~\cite{meynet_pc-msdm_2019}, PC-ASIM~\cite{alexiou_point_2018}, PCQM~\cite{9123147}, and PointSSIM~\cite{9106005}. The other is the projection-based metric~\cite{torlig_novel_2018}. In the proposed projection-based PCQA methods and the benchmark scheme~\cite{torlig_novel_2018}, 11 different 2D IQA submetrics were used to do the quality prediction, which are error based (NQM~\cite{nqm_841940}), structural similarity based (SSIM~\cite{1284395}, MS-SSIM~\cite{1292216}, IW-SSIM~\cite{iwssim_5635337}, 
GSM~\cite{gsm_6081939}, GMSD~\cite{gmsd_6678238}, 
RFSIM~\cite{5649275}, SR-SIM~\cite{6467149}, VSI~\cite{vsi_6873260}), and natural scene statistics based (IFC~\cite{ifc_1532311}, VIF~\cite{1576816}). 
PLCC, SROCC, KORCC and RMSE~\cite{itu1401} were used to evaluate the performance of different PCQA metrics.

\begin{table*}
\caption{Performance of point-based and projection-based objective point cloud quality metrics.}
\begin{center}
\resizebox{\textwidth}{!}{
\begin{tabular}{c|l|c c c c|c c c c | c c c c }
\hline
\multirow{3}{*}{\tabincell{l}{Point-based and \\Projection-based\\PCQA methods}}
& \multirow{3}{*}{Submetric} & \multicolumn{4}{c|}{\multirow{2}{*}{All}} & \multicolumn{4}{c|}{\multirow{2}{*}{Human figure session}} & \multicolumn{4}{c}{\multirow{2}{*}{Inanimate object session}} \\ 

&&&&&&&&&&&&&\\
 & & PLCC & SROCC & KROCC & RMSE & PLCC & SROCC & KROCC & RMSE & PLCC & SROCC & KROCC & RMSE \\ 
\hline

PC-MSDM~\cite{meynet_pc-msdm_2019}& \multirow{4}{*}{-}&0.1814&0.1470&0.0991&0.1293&0.3003&0.2981&0.2125&0.1289&0.1902&0.0024&0.0008&0.1253\\
PC-ASIM~\cite{alexiou_point_2018}&  &0.2374&0.2695&0.1780&0.1277&0.3526&0.3382&0.2228&0.1265&0.1375&0.2389&0.1660&0.1264 \\
PCQM~\cite{9123147}& &0.6539&0.6666&0.4825&0.0994&0.6391&\underline{0.6538}&\underline{0.4757}&0.1040&0.6811&0.6882&0.5069&0.0934 \\
PointSSIM~\cite{9106005}&&\underline{0.7808}&\underline{0.6955}&\underline{0.5086}&\underline{0.0821}&\underline{0.7294}&0.6152&0.4449&\underline{0.0925}&\underline{0.8785}&\underline{0.8033}&\underline{0.6180}&\underline{0.0610} \\\hline

\multirow{13}{*}{\tabincell{l}{View Projection Based~\cite{torlig_novel_2018}}} 

&GMSD  &
0.3487&0.2713&0.1874&0.1208&0.4511&0.3034&0.2148&0.1181&0.4172&0.3555&0.2502&0.1140 

 \\
&GSM & 
0.1447&0.1828&0.1277&0.1276&0.3645&0.1993&0.1387&0.1232&0.2586&0.2476&0.1723&0.1212 

\\ 
 & IFC &
\underline{0.5378}&0.5123&0.3528&\underline{0.1087}&\underline{0.6669}&\underline{0.6126}&\underline{0.4360}&\underline{0.0986}&0.5477&0.5635&0.3979&0.1050 

  \\
&IW-SSIM & 
0.4298&0.4044&0.2714&0.1164&0.5620&0.5453&0.3723&0.1095&0.3684&0.3368&0.2228&0.1166 

\\
&MS-SSIM & 
0.1647&0.2302&0.1575&0.1272&0.4247&0.2867&0.2042&0.1198&0.2570&0.2625&0.1831&0.1212 

\\
&NQM & 
0.2014&0.2357&0.1629&0.1263&0.2658&0.2998&0.1977&0.1276&0.2512&0.2992&0.2070&0.1214 


\\
&RFSIM&
0.2269&0.2352&0.1550&0.1256&0.3315&0.3715&0.2364&0.1248&0.3167&0.3145&0.2158&0.1190 

\\
&SR-SIM&
0.2405&0.2888&0.1960&0.1251&0.4114&0.3491&0.2342&0.1206&0.4560&0.4408&0.2966&0.1117 

\\ 
&SSIM & 
0.1027&0.1344&0.0938&0.1283&0.0204&0.1330&0.0935&0.1323&0.1793&0.2295&0.1566&0.1234

\\ 
&VIF & 
0.5284&\underline{0.5564}&\underline{0.3836}&0.1095&0.5580&0.5549&0.3914&0.1098&\underline{0.5862}&\underline{0.6324}&\underline{0.4382}&\underline{0.1016} 

\\
&VSI & 
0.3445&0.2920&0.1985&0.1210&0.4649&0.4144&0.2874&0.1172&0.3791&0.3751&0.2536&0.1161 

\\
\hline 
\multirow{13}{*}{\tabincell{l}{\textbf{Proposed Weighted}\\\textbf{View Projection Based}}}
&GMSD  &
0.3959&0.3069&0.2103&0.1184&0.3936&0.3298&0.2321&0.1216&0.5012&0.4105&0.2826&0.1085 

 \\
&GSM & 
0.2135&0.2122&0.1469&0.1260&0.3647&0.2213&0.1556&0.1232&0.3346&0.3161&0.2150&0.1182 

\\ 
 & IFC &
0.4941&0.4657&0.3154&0.1121&\underline{0.6716}&\underline{0.6211}&\underline{0.4497}&\underline{0.0980}&0.4411&0.4236&0.2806&0.1126 

  \\
&IW-SSIM & 
0.4394&0.4112&0.2784&0.1158&0.5480&0.5356&0.3653&0.1107&0.3871&0.3386&0.2212&0.1157 

\\
&MS-SSIM & 
0.3301&0.2585&0.1777&0.1217&0.4399&0.3020&0.2183&0.1188&0.3834&0.3214&0.2249&0.1159 

\\
&NQM & 
0.2209&0.2312&0.1601&0.1257&0.3120&0.2773&0.1848&0.1257&0.2623&0.3015&0.2060&0.1211 


\\
&RFSIM&
0.2278&0.2415&0.1616&0.1255&0.3265&0.3673&0.2372&0.1251&0.3486&0.3056&0.2109&0.1176 

\\
&SR-SIM&
0.3308&0.2921&0.1983&0.1217&0.4235&0.3439&0.2335&0.1199&0.4216&0.4308&0.2872&0.1138

\\ 
&SSIM & 
0.2294&0.1715&0.1205&0.1255&0.3501&0.1507&0.1063&0.1240&0.3601&0.3248&0.2210&0.1170

\\ 
&VIF & 
\underline{0.5474}&\underline{0.5595}&\underline{0.3852}&\underline{0.1079}&0.5630&0.5594&0.3967&0.1094&\underline{0.6106}&\underline{0.6210}&\underline{0.4277}&\underline{0.0993} 

\\
&VSI & 
0.3315&0.2922&0.1993&0.1216&0.4416&0.4110&0.2865&0.1187&0.3837&0.3434&0.2312&0.1158 

\\
\cline{2-14}
&\textbf{Average gain}&
\textbf{0.0446}&\textbf{0.0090}&\textbf{0.0061}&\textbf{-0.0013}&\textbf{0.0285}&\textbf{0.0045}&\textbf{0.0054}&\textbf{-0.0006}&\textbf{0.0379}&\textbf{0.0072}&\textbf{0.0013}&\textbf{-0.0014}
 \\
&\textbf{Ratio} & 
\textbf{14.31\%}&\textbf{2.92\%}&\textbf{2.88\%}&\textbf{-1.10\%}&\textbf{6.70\%}&\textbf{1.26\%}&\textbf{2.19\%}&\textbf{-0.49\%}&\textbf{10.12\%}&\textbf{1.96\%}&\textbf{0.50\%}&\textbf{-1.26\%}
 \\

\hline 
\multirow{13}{*}{\tabincell{l}{\textbf{Proposed}\\\textbf{Patch Projection Based}}}

&GMSD&
0.7360&0.5923&0.4208&0.0873&0.7993&0.6169&0.4476&0.0795&0.6992&0.5947&0.4257&0.0897 
 \\
&GSM&
0.6536&0.5408&0.3777&0.0976&0.7968&0.6056&0.4416&0.0800&0.5394&0.4592&0.3225&0.1056 

\\
&IFC&
0.2925&0.2404&0.1664&0.1233&0.3440&0.1353&0.0936&0.1243&0.3400&0.3824&0.2785&0.1180 

 \\
&IW-SSIM&
\underline{0.8181}&\underline{0.6966}&\underline{0.5183}&\underline{0.0742}&\underline{0.8101}&0.6505&0.4761&\underline{0.0776}&\underline{0.8994}&\underline{0.8563}&\underline{0.6783}&\underline{0.0548} 
\\
&MS-SSIM&
0.6772&0.5180&0.3632&0.0949&0.7796&0.5369&0.3827&0.0829&0.5945&0.4924&0.3482&0.1009 
\\
&NQM&
0.5100&0.5082&0.3523&0.1109&0.7855&0.6783&\underline{0.5119}&0.0819&0.5172&0.5547&0.3979&0.1074

\\
&RFSIM&
 
0.6144&0.5493&0.3798&0.1017&0.7636&0.6345&0.4535&0.0854&0.5841&0.5281&0.3716&0.1018 

\\
&SR-SIM&
0.7761&0.6539&0.4731&0.0813&0.8040&\underline{0.6527}&0.4796&0.0787&0.8349&0.7748&0.5816&0.0690 

\\
&SSIM&
0.5498&0.4280&0.2915&0.1077&0.6259&0.4521&0.3116&0.1032&0.5436&0.4257&0.2997&0.1053

\\
&VIF&
0.6043&0.5404&0.3751&0.1027&0.6914&0.5757&0.4025&0.0956&0.7749&0.6925&0.5225&0.0793 
 
\\
&VSI&
0.8063&0.6807&0.5013&0.0763&0.7994&0.6359&0.4626&0.0795&0.8922&0.8429&0.6540&0.0567 

\\

\cline{2-14}
&\textbf{Average gain}&
\textbf{0.3426}&\textbf{0.2368}&\textbf{0.1757}&\textbf{-0.0253}&\textbf{0.3162}&\textbf{0.1912}&\textbf{0.1504}&\textbf{-0.0303}&\textbf{0.2912}&\textbf{0.2316}&\textbf{0.1897}&\textbf{-0.0257} 

 \\
&\textbf{Ratio} & 
\textbf{109.94\%}&\textbf{77.07\%}&\textbf{82.90\%}&\textbf{-21.22\%}&\textbf{74.42\%}&\textbf{53.58\%}&\textbf{60.91\%}&\textbf{-25.82\%}&\textbf{77.72\%}&\textbf{62.56\%}&\textbf{73.53\%}&\textbf{-22.79\%}

 \\

\hline

\end{tabular}
}
\end{center}
\label{tbl:performance_2D_projection}
\end{table*}

\subsection{Evaluation for View Projection Based PCQA}
Table~\ref{tbl:performance_2D_projection} shows the correlation between subjective scores and objective scores of different projection-based PCQA methods. 
An average gain $\rho_G$ between the proposed scheme and the benchmark scheme is computed as 
\begin{equation}
    \rho_G =  \frac{1}{N} \sum_{i=1}^N {(X_i-X_{org_i})},
\end{equation}
and the gain ratio $\rho_R$ is evaluated as 
\begin{equation}
    \rho_R = \frac{\sum_{i=1}^N{(X_i-X_{org_i})}}{\sum_{i=1}^N{X_{org_i}}},
\end{equation}
where $X_{org_i}$ and $X_i$ denote the PLCC, SROCC, KORCC and RMSE of the PCQA~\cite{torlig_novel_2018} and the proposed PCQA using submetrics $i$. $N=11$ since 11 submetrics were tested. 

Table~\ref{tbl:performance_2D_projection} shows the PLCC, SROCC, KORCC and RMSE comparison between the proposed weighted view projection based PCQA and the benchmark schemes. In the table, the bold denotes the average gain and the gain ratio of the proposed method, and the underlined denotes the best performance in each scheme. 
We have two key observations. 
1) Compared with the view projection based method~\cite{torlig_novel_2018}, the proposed weighted view projection based method is able to achieve an increase of 0.0285, 0.0379, and 0.0446 in PLCC on average (6.70\%, 10.12\%, 14.31\%) for Human figure subset, Inanimate object subset, and ALL in SIAT-PCQD, respectively. Similarly, gains can be found for SROCC, KROCC, and RMSE, which proves the proposed weighted view projection based method is more effective than the benchmark~\cite{torlig_novel_2018}. 
2) While comparing the effectiveness of using different submetrics, the proposed weighted view projection based method achieves the best when using IFC~\cite{ifc_1532311} and VIF~\cite{1576816}.

\subsection{Evaluation for Patch Projection Based PCQA}
The bottom row of Table~\ref{tbl:performance_2D_projection} shows the PLCC, SROCC, KORCC and RMSE of the proposed patch projection based PCQA. 
We have four key observations. 
1) Compared with the view projection based method~\cite{torlig_novel_2018}, the proposed patch projection based method is able to achieve an increase of 0.3162, 0.2912, and 0.3426 in PLCC on average (74.42\%, 77.72\%, 109.94\%) for Human figure subset, Inanimate object subset, and ALL in SIAT-PCQD, respectively. Similarly, gains can be found for SROCC, KROCC, and RMSE, which are significantly higher than those of the proposed weighted view projection based method. 
2) While comparing the effectiveness of using different submetrics, the proposed patch projection based method achieves the best when using IW-SSIM~\cite{iwssim_5635337} whose PLCC is 0.8101, 0.8994, and 0.8181 for subsets and the whole SIAT-PCQD database.
3) As for the point-based methods, the PLCCs of PC-MSDM~\cite{meynet_pc-msdm_2019}, PC-ASIM~\cite{alexiou_point_2018} PCQM~\cite{9123147}, and PointSSIM~\cite{9106005} are 0.1814, 0.2374, 0.6539, and 0.7808 in SIAT-PCQD, respectively. Compared with them, the proposed patch projection based method using IW-SSIM and VSI achieves a higher PLCC as 0.8181 and 0.8063, respectively. 
4) Compared with the D1, D2, and YUV in Table~\ref{tbl:performance_3D}, the proposed patch projection based method outperforms significantly these schemes in PLCC, SROCC, KROCC, and RMSE in SIAT-PCQD.

To further validate the effectiveness of the proposed patch projection based PCQA while comparing with the view projection based scheme~\cite{torlig_novel_2018}, another point cloud database, named vsenseVVDB~\cite{zerman_subjective_2019}, was tested and Table~\ref{tbl:performance_patch_vsense} shows the results. 
We can observe 1) the average gain and the gain ratio are 0.4478 and 112.78\% in PLCC, respectively. Similarly, gains can be found for SROCC, KROCC, and RMSE. 
2) While using different submetrics in the proposed patch projection based method, the PLCCs range from 0.5545 to 0.9477. SR-SIM and GSM achieve the top two performance in PLCC. 
The proposed patch projection based PCQA is significantly better than the view projection based scheme in vsenseVVDB. 

The comparative studies on seven PCQA methods and two databases illustrate the proposed patch projection based method is more effective. The advantage of the proposed patch projection based method is it relieves the occlusion problem in view projection by segmenting the point cloud into smaller parts. Secondly, the framework of the proposed patch projection based method is able to take advantage of the advanced 2D IQA metrics for PCQA.

\begin{table}
    \caption{Performance of the proposed patch projection based obejctive method in the vsenseVVDB~\cite{zerman_subjective_2019} database }
    \begin{center}
    \resizebox{\linewidth}{!}{
    \begin{tabular}{c|l|c c c c}
    \hline
    \tabincell{l}{Projection-based\\PCQA methods}&Submetric& PLCC & SROCC & KROCC & RMSE  \\ 
    \hline
    
    \multirow{11}{*}{
        \tabincell{l}{View Projection Based~\cite{torlig_novel_2018}}
        }
    &GMSD&0.2529&0.2813&0.1984&24.2996\\
    &GSM&0.3468&0.3484&0.2452&23.5575\\
    &IFC&0.6891&0.6799&0.4777&18.2002\\
    &IW-SSIM&0.469&0.4485&0.3239&22.183\\
    &MS-SSIM&0.2549&0.2831&0.1943&24.9611\\
    &NQM&0.5382&0.2666&0.1943&21.1682\\
    &RFSIM&0.4416&0.4415&0.3279&22.5341\\
    &SR-SIM&0.2319&0.3333&0.2308&24.4312\\
    &SSIM&0.2396&0.275&0.1984&24.3844\\
    &VIF&0.4885&0.4653&0.3279&21.9154\\
    &VSI&0.4154&0.2692&0.1903&24.5985\\
    \hline
    \multirow{13}{*}{
        \tabincell{c}{\textbf{Proposed}\\\textbf{Patch Projection Based}}
        }
    &GMSD&0.7101&0.824&0.6316&17.6826\\
    &GSM&0.9383&0.8939&0.7523&8.6856\\
    &IFC&0.8732&0.8346&0.6518&12.2409\\
    &IW-SSIM&0.8998&0.8254&0.6316&10.9596\\
    &MS-SSIM&0.8941&0.8225&0.6342&11.2509\\
    &NQM&0.8273&0.7983&0.587&14.108\\
    &RFSIM&0.9011&\underline{0.9322}&0.7611&23.8582\\
    &SR-SIM&\underline{0.9477}&0.8951&0.7264&\underline{8.0182}\\
    &SSIM&0.5545&0.5907&0.4228&20.9014\\
    &VIF&0.8389&0.8254&0.6437&13.6709\\
    &VSI&0.9094&0.9051&\underline{0.7658}&10.4448\\
    \cline{2-6}
    &\textbf{Average gain}&\textbf{0.4478}&\textbf{0.4596}&\textbf{0.3908}&\textbf{-9.1284}\\
    &\textbf{Ratio}&\textbf{112.78\%}&\textbf{123.54\%}&\textbf{147.78\%}&\textbf{-39.81\%}\\
    \hline
    \end{tabular}
    }
    \end{center}
    \label{tbl:performance_patch_vsense}
\end{table}

\section{Conclusions}\label{section7}
In this paper, a subjective point cloud quality assessment experiment in an immersive virtual reality environment with a head-mounted display was conducted, and a weighted view projection based objective method and a patch projection based objective method were proposed. 
The impacts of sequences and geometry and texture quantization parameters were discussed in the analyses of the subjective experiment. 
The proposed patch projection based method improved the correlation of predictive scores with subjective scores based on image quality assessment metrics for the reason that the method reduces occluded areas during the process of projection. 
Nowadays, point cloud assessment is still an intricate and challenging problem involved with excessive elements. Our subjective database and findings can be used in perception-based point cloud processing, transmission, and coding, especially for virtual reality applications.

\bibliographystyle{IEEEtran} 
\bibliography{SIAT_PCQD}

\begin{thebibliography}{10}
\providecommand{\url}[1]{#1}
\csname url@samestyle\endcsname
\providecommand{\newblock}{\relax}
\providecommand{\bibinfo}[2]{#2}
\providecommand{\BIBentrySTDinterwordspacing}{\spaceskip=0pt\relax}
\providecommand{\BIBentryALTinterwordstretchfactor}{4}
\providecommand{\BIBentryALTinterwordspacing}{\spaceskip=\fontdimen2\font plus
\BIBentryALTinterwordstretchfactor\fontdimen3\font minus
  \fontdimen4\font\relax}
\providecommand{\BIBforeignlanguage}[2]{{%
\expandafter\ifx\csname l@#1\endcsname\relax
\typeout{** WARNING: IEEEtran.bst: No hyphenation pattern has been}%
\typeout{** loaded for the language `#1'. Using the pattern for}%
\typeout{** the default language instead.}%
\else
\language=\csname l@#1\endcsname
\fi
#2}}
\providecommand{\BIBdecl}{\relax}
\BIBdecl

\bibitem{schwarz_emerging_2019}
S.~{Schwarz}, M.~{Preda}, V.~{Baroncini}, M.~{Budagavi}, P.~{Cesar}, P.~A.
  {Chou}, R.~A. {Cohen}, M.~{Krivokuća}, S.~{Lasserre}, Z.~{Li}, J.~{Llach},
  K.~{Mammou}, R.~{Mekuria}, O.~{Nakagami}, E.~{Siahaan}, A.~{Tabatabai}, A.~M.
  {Tourapis}, and V.~{Zakharchenko}, ``Emerging mpeg standards for point cloud
  compression,'' \emph{IEEE Trans. Emerg. Sel. Topics Circuits Syst.}, vol.~9,
  no.~1, pp. 133--148, March 2019.

\bibitem{mekuria_design_2016}
R.~Mekuria, K.~Blom, and P.~Cesar, ``Design, implementation, and evaluation of
  a point cloud codec for tele-immersive video,'' \emph{IEEE Trans. Circuits
  Syst. Video Technol.}, vol.~27, no.~4, pp. 828--842, 2016.

\bibitem{alexiou_towards_2017}
E.~Alexiou, E.~Upenik, and T.~Ebrahimi, ``Towards subjective quality assessment
  of point cloud imaging in augmented reality,'' in \emph{2017 {IEEE} 19th
  {Int.} {Workshop} {Multimedia} {Signal} {Process.} ({MMSP})}, 2017, pp. 1--6.

\bibitem{javaheri_subjective_2017}
A.~Javaheri, C.~Brites, F.~Pereira, and J.~Ascenso, ``Subjective and objective
  quality evaluation of {3D} point cloud denoising algorithms,'' in \emph{2017
  {IEEE} {Int.} {Conf.} {Multimedia} {Expo} {Workshops} ({ICMEW})}, 2017, pp.
  1--6.

\bibitem{alexiou_performance_2017}
E.~Alexiou and T.~Ebrahimi, ``On the performance of metrics to predict quality
  in point cloud representations,'' in \emph{Appl. {Digit.} {Image} {Process.}
  {XL}}, vol. 10396, 2017, p. 103961H.

\bibitem{alexiou_impact_2018}
{E. Alexiou and T. Ebrahimi}, ``Impact of {Visualisation} {Strategy} for
  {Subjective} {Quality} {Assessment} of {Point} {Clouds},'' in \emph{2018
  {IEEE} {Int.} {Conf.} {Multimedia} {Expo} {Workshops} ({ICMEW})}, 2018, pp.
  1--6.

\bibitem{alexious_point_2018}
E.~Alexiou, A.~M. Pinheiro, C.~Duarte, D.~Matković, E.~Dumić, L.~A.
  da~Silva~Cruz, L.~G. Dmitrović, M.~V. Bernardo, M.~Pereira, and T.~Ebrahimi,
  ``Point cloud subjective evaluation methodology based on reconstructed
  surfaces,'' in \emph{Appl. {Digit.} {Image} {Process.} {XLI}}, vol. 10752,
  2018, p. 107520H.

\bibitem{alexiou_point_2018-2}
E.~Alexiou, T.~Ebrahimi, M.~V. Bernardo, M.~Pereira, A.~Pinheiro, L.~A. D.~S.
  Cruz, C.~Duarte, L.~G. Dmitrovic, E.~Dumic, D.~Matkovics, and {others},
  ``Point cloud subjective evaluation methodology based on {2D} rendering,'' in
  \emph{2018 {10th} {Int.} {Conf.} {Quality} {Multimedia} {Experience}
  ({QoMEX})}, 2018, pp. 1--6.

\bibitem{siat-pcqd}
\BIBentryALTinterwordspacing
X.~Wu, Y.~Zhang, C.~Fan, J.~Hou, and S.~Kwong, ``Siat-pcqd: Subjective point
  cloud quality database with 6dof head-mounted display,'' 2021. [Online].
  Available: \url{https://dx.doi.org/10.21227/ad8d-7r28}
\BIBentrySTDinterwordspacing

\bibitem{torlig_novel_2018}
E.~M. Torlig, E.~Alexiou, T.~A. Fonseca, R.~L. de~Queiroz, and T.~Ebrahimi, ``A
  novel methodology for quality assessment of voxelized point clouds,'' in
  \emph{Appl. {Digit.} {Image} {Process.} {XLI}}, vol. 10752, 2018, p. 107520I.

\bibitem{alexiou_comprehensive_2019}
E.~Alexiou, I.~Viola, T.~M. Borges, T.~A. Fonseca, R.~L. de~Queiroz, and
  T.~Ebrahimi, ``A comprehensive study of the rate-distortion performance in
  mpeg point cloud compression,'' \emph{APSIPA Trans. Signal Inf. Process.},
  vol.~8, p. e27, 2019.

\bibitem{zhang_subjective_2014}
J.~Zhang, W.~Huang, X.~Zhu, and J.-N. Hwang, ``A subjective quality evaluation
  for {3D} point cloud models,'' in \emph{2014 {Int.} {Conf.} {Audio},
  {Language} and {Image} {Process.}}, 2014, pp. 827--831.

\bibitem{javaheri_subjective_2017-1}
A.~Javaheri, C.~Brites, F.~Pereira, and J.~Ascenso, ``Subjective and objective
  quality evaluation of compressed point clouds,'' in \emph{2017 {IEEE} 19th
  {Int.} {Workshop} {Multimedia} {Signal} {Process.} ({MMSP})}, 2017, pp. 1--6.

\bibitem{da_silva_cruz_point_2019}
L.~A. da~Silva~Cruz, E.~Dumić, E.~Alexiou, J.~Prazeres, R.~Duarte, M.~Pereira,
  A.~Pinheiro, and T.~Ebrahimi, ``Point cloud quality evaluation: {Towards} a
  definition for test conditions,'' in \emph{2019 {11th} {Int.} {Conf.}
  {Quality} {Multimedia} {Experience} ({QoMEX})}, 2019, pp. 1--6.

\bibitem{9238424}
Q.~{Yang}, H.~{Chen}, Z.~{Ma}, Y.~{Xu}, R.~{Tang}, and J.~{Sun}, ``Predicting
  the perceptual quality of point cloud: A 3d-to-2d projection-based
  exploration,'' \emph{IEEE Trans. Multimedia}, pp. 1--1, 2020.

\bibitem{zerman_subjective_2019}
E.~Zerman, P.~Gao, C.~Ozcinar, and A.~Smolic, ``Subjective and objective
  quality assessment for volumetric video compression,'' \emph{Electron.
  Imag.}, vol. 2019, no.~10, pp. 323--1, 2019.

\bibitem{su_perceptual_2019}
H.~Su, Z.~Duanmu, W.~Liu, Q.~Liu, and Z.~Wang, ``Perceptual {Quality}
  {Assessment} of 3d {Point} {Clouds},'' in \emph{2019 {IEEE} {Int.} {Conf.}
  {Image} {Process.} ({ICIP})}, 2019, pp. 3182--3186.

\bibitem{javaheri_point_2019}
A.~Javaheri, C.~Brites, F.~Pereira, and J.~Ascenso, ``Point {Cloud} {Rendering}
  after {Coding}: {Impacts} on {Subjective} and {Objective} {Quality},''
  \emph{arXiv preprint arXiv:1912.09137}, 2019.

\bibitem{9123137}
E.~{Zerman}, C.~{Ozcinar}, P.~{Gao}, and A.~{Smolic}, ``Textured mesh vs
  coloured point cloud: A subjective study for volumetric video compression,''
  in \emph{2020 Int. Conf. Quality Multimedia Experience (QoMEX)}, 2020, pp.
  1--6.

\bibitem{9200318}
K.~{Cao}, Y.~{Xu}, and P.~{Cosman}, ``Visual quality of compressed mesh and
  point cloud sequences,'' \emph{IEEE Access}, vol.~8, pp. 171\,203--171\,217,
  2020.

\bibitem{9191308}
S.~{Perry}, H.~P. {Cong}, L.~A. {da Silva Cruz}, J.~{Prazeres}, M.~{Pereira},
  A.~{Pinheiro}, E.~{Dumic}, E.~{Alexiou}, and T.~{Ebrahimi}, ``Quality
  evaluation of static point clouds encoded using mpeg codecs,'' in \emph{2020
  {IEEE} {Int.} {Conf.} {Image} {Process.} ({ICIP})}, 2020, pp. 3428--3432.

\bibitem{9089539}
S.~{Subramanyam}, J.~{Li}, I.~{Viola}, and P.~{Cesar}, ``Comparing the quality
  of highly realistic digital humans in 3dof and 6dof: A volumetric video case
  study,'' in \emph{2020 IEEE Conf. Virtual Reality and 3D User Interfaces
  (VR)}, 2020, pp. 127--136.

\bibitem{9123121}
E.~{Alexiou}, N.~{Yang}, and T.~{Ebrahimi}, ``Pointxr: A toolbox for
  visualization and subjective evaluation of point clouds in virtual reality,''
  in \emph{2020 Int. Conf. Quality Multimedia Experience (QoMEX)}, 2020, pp.
  1--6.

\bibitem{N17995}
{S. Schwarz and D. Flynn}, ``Common test conditions for point cloud
  compression,'' ISO/IEC JTC1/SC29/WG11, Macau, Tech. Rep. N17995, October
  2018.

\bibitem{tian_geometric_2017}
D.~Tian, H.~Ochimizu, C.~Feng, R.~Cohen, and A.~Vetro, ``Geometric distortion
  metrics for point cloud compression,'' in \emph{2017 {IEEE} {Int.} {Conf.}
  {Image} {Process.} ({ICIP})}, 2017, pp. 3460--3464.

\bibitem{alexiou_point_2018}
E.~Alexiou and T.~Ebrahimi, ``Point cloud quality assessment metric based on
  angular similarity,'' in \emph{2018 {IEEE} {Int.} {Conf.} {Multimedia} {Expo}
  ({ICME})}, 2018, pp. 1--6.

\bibitem{meynet_pc-msdm_2019}
G.~Meynet, J.~Digne, and G.~Lavoué, ``{PC}-{MSDM}: {A} quality metric for {3D}
  point clouds,'' in \emph{2019 {11th} {Int.} {Conf.} {Quality} {Multimedia}
  {Experience} ({QoMEX})}, 2019, pp. 1--3.

\bibitem{9191233}
A.~{Javaheri}, C.~{Brites}, F.~{Pereira}, and J.~{Ascenso}, ``Improving
  psnr-based quality metrics performance for point cloud geometry,'' in
  \emph{2020 {IEEE} {Int.} {Conf.} {Image} {Process.} ({ICIP})}, 2020, pp.
  3438--3442.

\bibitem{9123087}
A.~Javaheri, C.~Brites, F.~Pereira, and J.~Ascenso, ``A generalized hausdorff
  distance based quality metric for point cloud geometry,'' in \emph{2020 Int.
  Conf. Quality Multimedia Experience (QoMEX)}, 2020, pp. 1--6.

\bibitem{9143408}
A.~{Javaheri}, C.~{Brites}, F.~{Pereira}, and J.~{Ascenso}, ``Mahalanobis based
  point to distribution metric for point cloud geometry quality evaluation,''
  \emph{IEEE Signal Process. Lett.}, vol.~27, pp. 1350--1354, 2020.

\bibitem{9190956}
R.~{Diniz}, P.~G. {Freitas}, and M.~C.~Q. {Farias}, ``Multi-distance point
  cloud quality assessment,'' in \emph{2020 {IEEE} {Int.} {Conf.} {Image}
  {Process.} ({ICIP})}, 2020, pp. 3443--3447.

\bibitem{9123076}
R.~Diniz, P.~G. Freitas, and M.~C.~Q. Farias, ``Towards a point cloud quality
  assessment model using local binary patterns,'' in \emph{2020 Int. Conf.
  Quality Multimedia Experience (QoMEX)}, 2020, pp. 1--6.

\bibitem{9123147}
G.~{Meynet}, Y.~{Nehmé}, J.~{Digne}, and G.~{Lavoué}, ``Pcqm: A
  full-reference quality metric for colored 3d point clouds,'' in \emph{2020
  Int. Conf. Quality Multimedia Experience (QoMEX)}, 2020, pp. 1--6.

\bibitem{9123089}
I.~{Viola}, S.~{Subramanyam}, and P.~{Cesar}, ``A color-based objective quality
  metric for point cloud contents,'' in \emph{2020 Int. Conf. Quality
  Multimedia Experience (QoMEX)}, 2020, pp. 1--6.

\bibitem{yang2020inferring}
Q.~Yang, Z.~Ma, Y.~Xu, Z.~Li, and J.~Sun, ``Inferring point cloud quality via
  graph similarity,'' \emph{arXiv preprint arXiv:2006.00497}, 2020.

\bibitem{9106005}
E.~{Alexiou} and T.~{Ebrahimi}, ``Towards a point cloud structural similarity
  metric,'' in \emph{2020 {IEEE} {Int.} {Conf.} {Multimedia} {Expo} {Workshops}
  ({ICMEW})}, 2020, pp. 1--6.

\bibitem{9198142}
I.~{Viola} and P.~{Cesar}, ``A reduced reference metric for visual quality
  evaluation of point cloud contents,'' \emph{IEEE Signal Process. Lett.},
  vol.~27, pp. 1660--1664, 2020.

\bibitem{alexiou_exploiting_2019}
E.~Alexiou and T.~Ebrahimi, ``Exploiting user interactivity in quality
  assessment of point cloud imaging,'' in \emph{2019 {11th} {Int.} {Conf.}
  {Quality} {Multimedia} {Experience} ({QoMEX})}, 2019, pp. 1--6.

\bibitem{seq:8i}
{E. d'Eon, B. Harrison, T. Myers, and P. A. Chou}, ``8i voxelized full bodies
  (a voxelized point cloud dataset),'' ISO/IEC JTC1/SC29/WG11, Geneva, Tech.
  Rep. m40059/M74006, January 2017.

\bibitem{seq:maria}
{T. Ebner, I. Feldmann, O. Schreer, P. Kauff, E. Feiler, F. Govaere, K.
  Costa-Zahn, and F. Mrongowius}, ``{HHI} point cloud dataset of moving
  actress,'' ISO/IEC JTC1/SC29/WG11, Gwangju, Korea, Tech. Rep. m42152, January
  2018.

\bibitem{seq:wegner}
{T. Ebner, I. Feldmann, O. Schreer, P. Kauff, and T. v. Unger}, ``{HHI} point
  cloud dataset of a boxing trainer,'' ISO/IEC JTC1/SC29/WG11, Ljubljana,
  Slovenia, Tech. Rep. m42921, July 2018.

\bibitem{seq:microsoft}
{C. Loop, Q. Cai, S. O. Escolano, and P. A. Chou}, ``Microsoft voxelized upper
  bodies - a voxelized point cloud dataset,'' Available at
  \url{https://jpeg.org/plenodb/pc/microsoft/} (accessed Apr. 4, 2020).

\bibitem{seq:roman}
{University of São Paulo}, ``Emerging image modalities representation and
  compression,'' Available at \url{http://uspaulopc.di.ubi.pt/} (accessed Apr.
  4, 2020).

\bibitem{seq:biplane}
EPFL, ``{ScanLAB} projects point cloud data sets,'' Available at
  \url{http://grebjpeg.epfl.ch/jpeg_pc/index_Bi-plane.html} (accessed Apr. 4,
  2020).

\bibitem{itu910}
{ITU-T}, ``Subjective video quality assessment methods for multimedia
  applications,'' {P.}, Tech. Rep. 910, 2008.

\bibitem{6280595}
S.~{Winkler}, ``Analysis of public image and video databases for quality
  assessment,'' \emph{IEEE J. Sel. Topics Signal Process.}, vol.~6, no.~6, pp.
  616--625, 2012.

\bibitem{itu500}
{ITU-R}, ``Methodology for the subjective assessment of the quality of
  television pictures,'' {BT.}, Tech. Rep. 500-13, 2012.

\bibitem{itu919}
{ITU-T}, ``Subjective test methodologies for 360° video on head-mounted
  displays,'' {P.}, Tech. Rep. 919, 2020.

\bibitem{8010398}
I.~{Viola}, M.~{Řeřábek}, and T.~{Ebrahimi}, ``Comparison and evaluation of
  light field image coding approaches,'' \emph{IEEE J. Sel. Topics Signal
  Process.}, vol.~11, no.~7, pp. 1092--1106, 2017.

\bibitem{perra2018assessing}
C.~Perra, ``Assessing the quality of experience in viewing rendered
  decompressed light fields,'' \emph{Multimedia Tools and Appl.}, vol.~77,
  no.~16, pp. 21\,771--21\,790, 2018.

\bibitem{1284395}
{Z. Wang}, A.~C. {Bovik}, H.~R. {Sheikh}, and E.~P. {Simoncelli}, ``Image
  quality assessment: from error visibility to structural similarity,''
  \emph{IEEE Trans. Image Process.}, vol.~13, no.~4, pp. 600--612, 2004.

\bibitem{5404314}
K.~{Seshadrinathan}, R.~{Soundararajan}, A.~C. {Bovik}, and L.~K. {Cormack},
  ``Study of subjective and objective quality assessment of video,'' \emph{IEEE
  Trans. Image Process.}, vol.~19, no.~6, pp. 1427--1441, 2010.

\bibitem{1576816}
H.~R. {Sheikh} and A.~C. {Bovik}, ``Image information and visual quality,''
  \emph{IEEE Trans. Image Process.}, vol.~15, no.~2, pp. 430--444, 2006.

\bibitem{itu1401}
{ITU-T}, ``Methods, metrics and procedures for statistical evaluation,
  qualification and comparison of objective quality prediction models,'' {P.},
  Tech. Rep. 1401, 2020.

\bibitem{9194311}
Q.~{Liu}, H.~{Yuan}, J.~{Hou}, R.~{Hamzaoui}, and H.~{Su}, ``Model-based joint
  bit allocation between geometry and color for video-based 3d point cloud
  compression,'' \emph{IEEE Trans. Multimedia}, pp. 1--1, 2020.

\bibitem{nqm_841940}
N.~{Damera-Venkata}, T.~D. {Kite}, W.~S. {Geisler}, B.~L. {Evans}, and A.~C.
  {Bovik}, ``Image quality assessment based on a degradation model,''
  \emph{IEEE Trans. Image Process.}, vol.~9, no.~4, pp. 636--650, 2000.

\bibitem{1292216}
Z.~{Wang}, E.~P. {Simoncelli}, and A.~C. {Bovik}, ``Multiscale structural
  similarity for image quality assessment,'' in \emph{37th Asilomar Conf.
  Signals, Syst. Computers, 2003}, vol.~2, 2003, pp. 1398--1402 Vol.2.

\bibitem{iwssim_5635337}
Z.~{Wang} and Q.~{Li}, ``Information content weighting for perceptual image
  quality assessment,'' \emph{IEEE Trans. Image Process.}, vol.~20, no.~5, pp.
  1185--1198, 2011.

\bibitem{gsm_6081939}
A.~{Liu}, W.~{Lin}, and M.~{Narwaria}, ``Image quality assessment based on
  gradient similarity,'' \emph{IEEE Trans. Image Process.}, vol.~21, no.~4, pp.
  1500--1512, 2012.

\bibitem{gmsd_6678238}
W.~{Xue}, L.~{Zhang}, X.~{Mou}, and A.~C. {Bovik}, ``Gradient magnitude
  similarity deviation: A highly efficient perceptual image quality index,''
  \emph{IEEE Trans. Image Process.}, vol.~23, no.~2, pp. 684--695, 2014.

\bibitem{5649275}
L.~{Zhang}, L.~{Zhang}, and X.~{Mou}, ``Rfsim: A feature based image quality
  assessment metric using riesz transforms,'' in \emph{2010 {IEEE} {Int.}
  {Conf.} {Image} {Process.} ({ICIP})}, 2010, pp. 321--324.

\bibitem{6467149}
L.~{Zhang} and H.~{Li}, ``Sr-sim: A fast and high performance iqa index based
  on spectral residual,'' in \emph{2012 {IEEE} {Int.} {Conf.} {Image}
  {Process.} ({ICIP})}, 2012, pp. 1473--1476.

\bibitem{vsi_6873260}
L.~{Zhang}, Y.~{Shen}, and H.~{Li}, ``Vsi: A visual saliency-induced index for
  perceptual image quality assessment,'' \emph{IEEE Trans. Image Process.},
  vol.~23, no.~10, pp. 4270--4281, 2014.

\bibitem{ifc_1532311}
H.~R. {Sheikh}, A.~C. {Bovik}, and G.~{de Veciana}, ``An information fidelity
  criterion for image quality assessment using natural scene statistics,''
  \emph{IEEE Trans. Image Process.}, vol.~14, no.~12, pp. 2117--2128, 2005.

\end{thebibliography}

%

\begin{IEEEbiography}[{\includegraphics[width=1in,height=1.25in,clip,keepaspectratio]{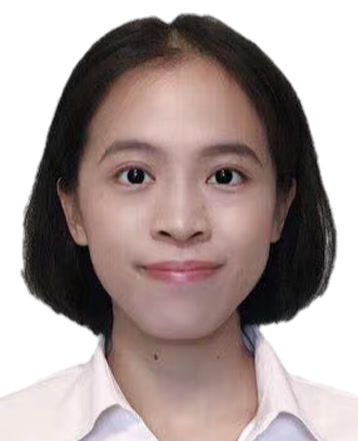}}]{Xinju Wu}
    received the B.S. degree in China University of Geosciences, Wuhuan, China, in 2018, and the M.S. degree with the Shenzhen Institute of Advanced Technology, University of Chinese Academy of Sciences, Shenzhen, China, in 2021. Her current research interests include point cloud quality assessment and processing.
\end{IEEEbiography}
    
\begin{IEEEbiography}[{\includegraphics[width=1in,height=1.25in,clip,keepaspectratio]{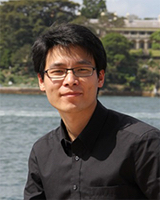}}]{Yun Zhang}
    received the B.S. and M.S. degrees in electrical engineering from Ningbo
    University, Ningbo, China, in 2004 and 2007, respectively, and the Ph.D. degree in computer science from Institute of Computing Technology (ICT), Chinese Academy of Sciences (CAS), Beijing, China, in 2010. From 2009 to 2014, he was a Visiting Scholar with the Department of Computer Science, City University of Hong Kong, Kowloon, Hong Kong. From 2010 to 2017, he was an Assistant Professor and an Associate Professor in Shenzhen Institute of Advanced Technology (SIAT), CAS, where he is currently a Professor in SIAT, CAS, Shenzhen, China. His research interests are multimedia communications and visual signal processing, computational visual perception, and machine learning.
\end{IEEEbiography}

\begin{IEEEbiography}[{\includegraphics[width=1in,height=1.25in,clip,keepaspectratio]{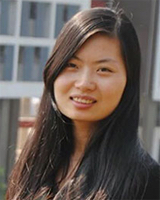}}]{Chunling Fan}
    received the M.S. degree from Nanjing Normal University, Nanjing, in 2011, and the Ph.D. degree with the Shenzhen Institute of Advanced Technology (SIAT), University of Chinese Academy of Sciences, Shenzhen, China, in 2020. She is now on the faculty of School of Electronic and Communication Engineering, Shenzhen Polytechnic, Shenzhen, China. Her research interests include image processing and image quality assessment.
\end{IEEEbiography}

\begin{IEEEbiography}[{\includegraphics[width=1in,height=1.25in,clip,keepaspectratio]{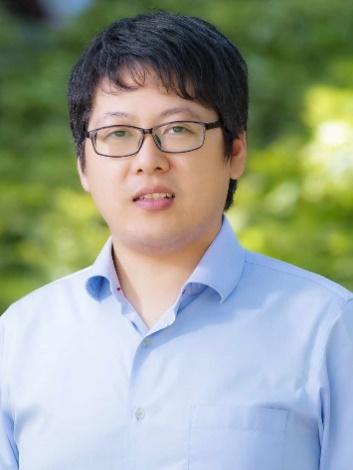}}]{Junhui Hou}
    received the B.Eng. degree in information engineering (Talented Students Program) from the South China University of Technology, Guangzhou, China, in 2009, the M.Eng. degree in signal and information processing from Northwestern Polytechnical University, Xian, China, in 2012, and the Ph.D. degree in electrical and electronic engineering from the School of Electrical and Electronic Engineering, Nanyang Technological University, Singapore, in 2016. He immediately joined the Department of Computer Science, City University of Hong Kong, as an Assistant Professor in Jan. 2017. His research interests fall into the general areas of visual computing, such as image/video/3D geometry data representation, processing and analysis, semi/un-supervised data modeling, and data compression and adaptive transmission

    Dr. Hou was the recipient of several prestigious awards, including the Chinese Government Award for Outstanding Students Study Abroad from China Scholarship Council in 2015 and the Early Career Award (3/381) from the Hong Kong Research Grants Council in 2018. He is an elected member of MSA-TC and VSPC-TC, IEEE CAS. He is currently an Associate Editor for IEEE Transactions on Circuits and Systems for Video Technology, IEEE Transactions on Image Processing, Signal Processing: Image Communication, and The Visual Computer. He also served as the Guest Editor for the IEEE Journal of Selected Topics in Applied Earth Observations and Remote Sensing and an Area Chair of ACM MM’19/20/21, IEEE ICME’20, VCIP’20/21, and WACV’21.  
\end{IEEEbiography}

\begin{IEEEbiography}[{\includegraphics[width=1in,height=1.25in,clip,keepaspectratio]{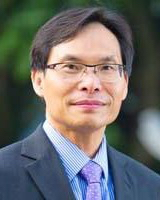}}]{Sam Kwong}
    received the B.S. and M.S. degrees in electrical engineering from State University of New York at Buffalo in 1983, University of Waterloo, Waterloo, ON, Canada, in 1985, and the Ph.D. degree from University of Hagen, Germany, in 1996. From 1985 to 1987, he was a Diagnostic Engineer with Control Data Canada. He joined Bell Northern Research Canada as a Member of Scientiﬁc Staff. In 1990, he became a Lecturer at the Department of Electronic Engineering, City University of Hong Kong, where he is currently a Professor at the Department of Computer Science, City University of Hong Kong, Kowloon, Hong Kong. His research interests are video and image coding and evolutionary algorithms.
\end{IEEEbiography}


\end{document}